\documentclass[acmsmall,screen,nonacm]{acmart}

\AtBeginDocument{%
  }

\setcopyright{cc}
\setcctype{by}
\acmJournal{PACMPL}
\acmYear{2025}
\acmVolume{9}
\acmNumber{PLDI}
\acmArticle{186}
\acmMonth{6}
\acmDOI{10.1145/3729289}

\newbool{extendedversion}
\setbool{extendedversion}{true}

\definecolor{lightgreen}{RGB}{66, 199, 30}
\definecolor{burgundy}{HTML}{530021}
\definecolor{tealgreen}{HTML}{198269}
\definecolor{rosered}{HTML}{D60B47}
\definecolor{mygrey}{HTML}{595959}
\usepackage{listings, listings-rust}
\newcommand{\rusti}[0]{\lstinline[language=Rust]}

\usepackage{mathpartir}
\usepackage{mathtools}

\usepackage{refable-rules}
\usepackage{rust-macros}
\usepackage{array}

\usepackage{tikz}
\usetikzlibrary{shapes.multipart, arrows, arrows.meta, calc, patterns}
\usepackage{xspace}
\usepackage{cleveref}
\usepackage{wrapfig}
\usepackage{enumitem}

\renewcommand{\sectionautorefname}{\S\kern-0.25em}
\renewcommand{\subsectionautorefname}{\S\kern-0.25em}

\begin{document}

\title{A Hybrid Approach to Semi-automated Rust Verification\ifbool{extendedversion}{ (Extended Version)}{}}


%
%
%

\author{Sacha-\'{E}lie Ayoun}
\orcid{0000-0001-9419-5387}
\email{s.ayoun17@imperial.ac.uk}
\affiliation{%
  \institution{Imperial College London}
  \city{London}
  \country{UK}
}

\author{Xavier Denis}
\orcid{0000-0003-2530-8418}
\email{research@xav.io}
\affiliation{%
  \institution{ETH Zurich}
  \city{Zurich}
  \country{Switzerland}
}

\author{Petar Maksimovi\'{c}}
\orcid{0000-0002-0400-7467}
\email{p.maksimovic@imperial.ac.uk}
\affiliation{%
  \institution{Nethermind}
  \city{London}
  \country{UK}
}
\affiliation{%
  \institution{Imperial College London}
  \city{London}
  \country{UK}
}

\author{Philippa Gardner}
\email{p.gardner@imperial.ac.uk}
\orcid{0000-0002-4187-0585}
\affiliation{%
  \institution{Imperial College London}
  \city{London}
  \country{UK}
}


\newtheorem{notation}{Notation}

\begin{abstract}
We propose a hybrid approach to end-to-end Rust verification where the proof effort is split into powerful automated verification of safe Rust and targeted semi-automated verification of unsafe~Rust. 
To this end, we present Gillian-Rust, a proof-of-concept semi-automated verification tool built on top of the Gillian platform that can reason about type safety and functional correctness of unsafe~code. Gillian-Rust automates a rich separation logic for real-world Rust, embedding the lifetime logic of RustBelt and the parametric prophecies of RustHornBelt, and is able to verify real-world Rust standard library code with only minor annotations and with verification times orders of magnitude faster than those of comparable tools.
We link Gillian-Rust with Creusot, a state-of-the-art verifier for safe Rust, by providing a systematic encoding of unsafe code specifications that Creusot can use but cannot verify, demonstrating the feasibility of our hybrid~approach.
\end{abstract}

\ccsdesc[300]{Theory of computation~Separation logic}
\ccsdesc[500]{Theory of computation~Automated reasoning}
\ccsdesc[500]{Theory of computation~Program verification}

\keywords{Rust, semi-automatic verification, symbolic execution, compositionality}


\maketitle


\vspace*{-0.2cm}
\section{Introduction}

Rust~\cite{rust-language-matsakis-2014, rust-programming-language-therustteam-} has seen rapid adoption in recent years, particularly in the field of \emph{systems programming}. 
Its success primarily stems from its rejection of false dichotomies between safety and performance: its \emph{ownership type system} and \emph{borrow checker} preserve memory safety while not needing garbage collection. 
With this success, however, also comes the need for stronger formal guarantees about the \emph{behaviour} of Rust programs, resulting in the development of tools such as Aeneas~\cite{aeneas-rust-verification-ho-2022}, Creusot~\cite{creusot-foundry-deductive-denis-2022} and Prusti~\cite{leveraging-rust-types-astrauskas-2019}. 
These tools all leverage the properties of the Rust type system to simplify verification, but all also share a common limitation: they can only verify \emph{safe} Rust code.

Real-world Rust code, however, commonly relies on \emph{unsafe} code to interface with the underlying operating system or provide low-level abstractions. Unsafe code gives the programmer `superpowers', such as the ability to dereference raw pointers, cast between types, and manipulate potentially uninitialised memory. It is an essential part of Rust's design, allowing for new \emph{safe} abstractions, such as \lstinline{LinkedList<T>} (the type of doubly-linked lists), to be implemented efficiently in libraries.
However, unsafe code also comes with greater responsibility: the onus is now on the programmer to ensure that their code does not exhibit undefined behaviour (UB) and that the corresponding APIs remain observationally safe. 
In addition, despite representing a fraction of the total codebase, unsafe code is often the most complex and error-prone part of a Rust program, making it the most important one to formally verify, which none of the above-mentioned tools is able to accomplish. 

We propose a \textbf{hybrid} approach to end-to-end Rust verification which, mirroring the differences between safe and unsafe code, leverages Creusot for verification of safe code and a novel tool, \emph{Gillian-Rust}, for verification of unsafe code, which can be specified but not verified by Creusot.

Understanding the substantial challenges that Gillian-Rust had to overcome requires in-depth knowledge of the related foundational work. 
In 2018, Jung et al. published RustBelt~\cite{rustbelt-securing-foundations-jung-2017}, a theoretical framework that allows for semantic interpretation of Rust ownership types using Iris~\cite{iris-ground-modular-jung-2018} and that can reason about type safety~(TS). 
In 2022, RustHornBelt~\cite{rusthornbelt-semantic-foundation-matsushita-2022} extended RustBelt with the ability to reason about functional correctness~(FC), allowing for safe functions implemented with unsafe code to be given first-order specifications and providing the meta-theory that now underpins Creusot.
Both RustBelt and RustHornBelt, however, work on $\lambdarust$, a model that makes simplifying assumptions expected of a foundational formalisation and does not capture the intricacies of real Rust. 
Moreover, RustHornBelt proofs are done manually in Rocq~\cite{coq-proof-assistant--}, on code ported by hand from Rust to $\lambdarust$, with little automation provided. More recently, RefinedRust~\cite{refined-rust} demonstrated how advanced automation techniques from Refined-C~\cite{refinedc-automating-foundational-sammler-2021} can be adapted to RustBelt to reason about FC of Rust programs.
However, RefinedRust remains embedded in Rocq, which inherently limits its automation and performance. We argue that more efficient and scalable tooling is needed in order for verification to tackle the volume of existing and future unsafe Rust code.

\myparagraph{Challenge 1: Tractable automated reasoning about the real-world Rust heap}
Real-world Rust comes with numerous systems-related complications, some known from C (e.g., low-level data representation and byte-level value manipulation) and some  new ones (e.g., zero-sized types, compiler-chosen layouts (C has a standardised layout), and polymorphism).
While these aspects remain invisible when using safe Rust, they become a proper concern when working with unsafe code. For example, a verifier must reason generically over all possible memory layouts of programs so that it could detect any potentially disallowed memory operations.
This makes reuse of existing memory models from C verification difficult and requires development of new techniques to reason automatically and efficiently about real-world Rust and the way it represents objects in~memory.

\myparagraph{Challenge 2: Type safety (TS), borrows, and raw pointers}
The notion of TS is much stricter in Rust than in languages like C. Specifically, the responsibility of a safe function, even an internally unsafe one, goes beyond its own body: it must ensure that no fully-safe program calling it may trigger UBs. This substantially increases the complexity of integrating unsafe code into a Rust~program.

The way Rust guarantees TS is through its strict and static ownership discipline, wherein each value must always have {\em an exclusive owner}. While this alone would be too restrictive, Rust also provides mutable references ($\mutref{\klft}\pty$) and shared references ($\&^{\klft}\pty$) which may \emph{borrow} ownership for a \emph{lifetime} $\klft$.
However, even when equipped with references, safe Rust is sometimes too restrictive and prevents the implementation of types such as \emph{doubly-linked lists}, where each node is referenced by two pointers at any time (cf. \autoref{fig:worlds}, bottom left), breaking exclusive ownership. In such cases, developers must resort to unsafe code in order to manipulate raw pointers ($*\mathtt{mut}~\pty$) which, unlike references, allow for unrestricted aliasing and do not provide any safety guarantees. This mixed use of raw pointers and safe references even further complicates the task of verifying TS of unsafe code, as it requires reasoning about lifetime-dependent safety invariants. 

\myparagraph{Challenge 3: Scaling safe and unsafe Rust verification, together}
While unsafe code is used to perform some of the most complex and primitive operations of Rust programs, it still comprises a small fraction of the total codebase~\cite{how-programmers-use-astrauskas-2020}.
Furthermore, safe Rust often uses many advanced features, such as higher-order functions, which are eschewed in unsafe code. 
For this reason, we believe that it would be highly challenging to build a tool that both has the required expressivity for reasoning about unsafe code, which makes extensive unrestricted use of raw pointers, and can, at the same time, reason \emph{efficiently} and \emph{automatically} about higher-level features used in safe~Rust.

On the other hand, tools such as Creusot~\cite{creusot-foundry-deductive-denis-2022, specifying-verifying-higherorder-denis-2023} have demonstrated that verification of {\em safe Rust only} can be performed with impressive automation and simplicity. Ideally, one would use such a tool for the safe part of a codebase, and another, more adapted tool, for analysing the unsafe part, dividing the proof effort appropriately. 
This approach, however, requires both tools to agree on the semantics of \emph{specifications} given to Rust functions. 
For example, if Creusot is used for safe code, the other tool has to provide a faithful interpretation of Creusot's specifications, which use a simple-to-write yet complex-to-interpret prophetic assertion language.

\begin{wrapfigure}{R}{0.45\textwidth}
\label{fig:main-diagram}
	\vspace*{-0.4cm}
    \centering
    \includegraphics[width=0.45\textwidth]{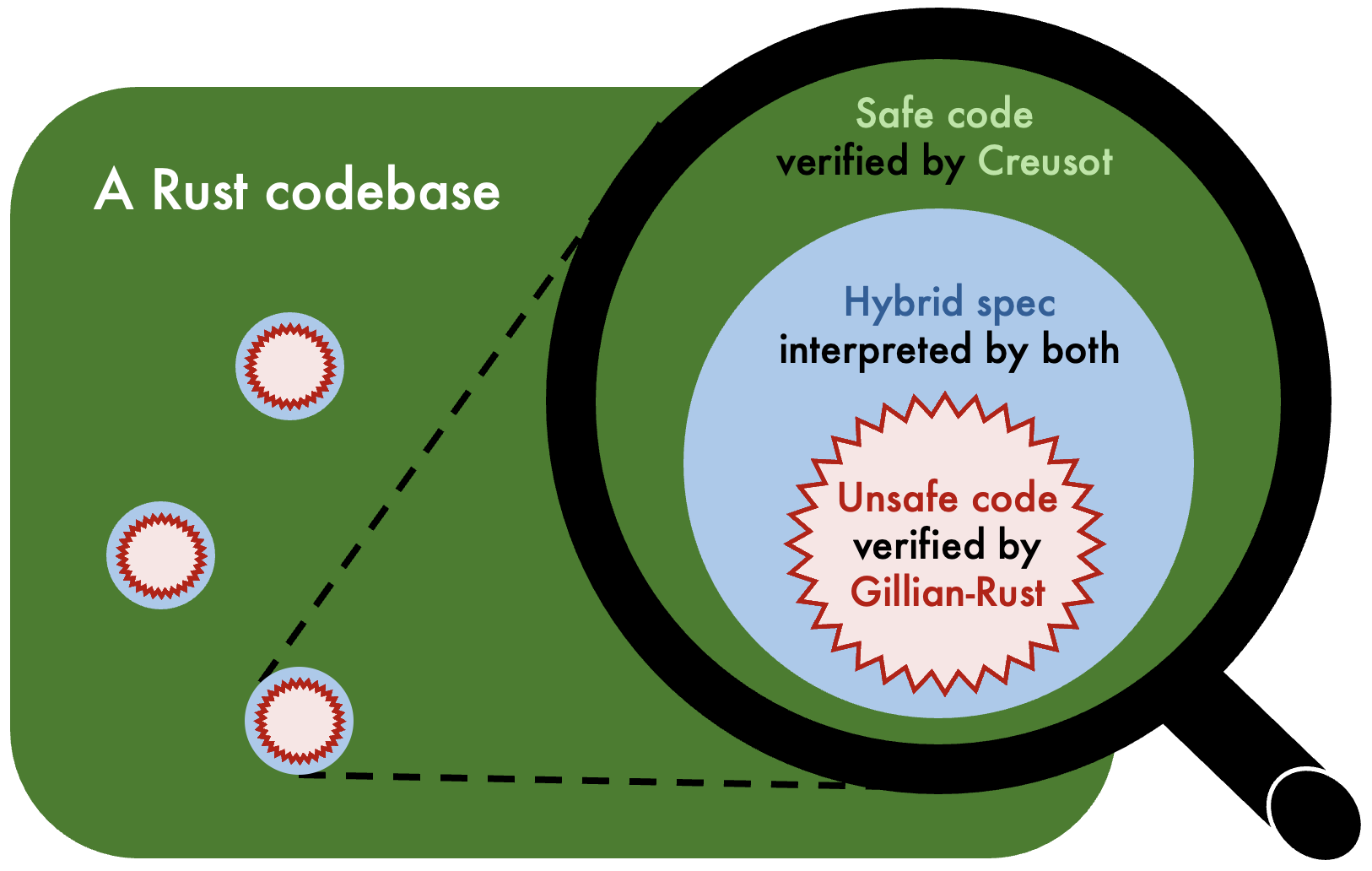}
    \vspace*{-0.8cm}
\end{wrapfigure}

\myparagraph{Contributions and paper outline} We present a \emph{hybrid} approach to Rust verification, which leverages the strengths of specialised tools operating in unison to verify both safe and unsafe Rust code, illustrated in the diagram on the right-hand side.

In particular, we combine Creusot, an existing tool for safe Rust verification, with \creusillian{}, a novel proof-of-concept semi-automated verification tool for unsafe Rust, built on top of the Gillian compositional symbolic analysis platform~\cite{gillian-part-ii-maksimovic-2021}.
We manage the boundary between the tools through a shared specification language that can easily be interpreted into either Creusot or Gillian-Rust specifications.
To make this possible, \creusillian{} implements and automates the reasoning of RustBelt and RustHornBelt, which allows it to reason about the \emph{prophetic specifications} of Creusot. 

We demonstrate the viability of our approach by verifying \emph{actual} Rust standard library code (specifically, the \rusti{LinkedList} and \rusti{Vec} types), along with several other case studies.
Our approach performs verification at least two orders of magnitude faster than prior works, made possible by the use of symbolic execution and the efficient memory model of Gillian-Rust.

The paper is structured as follows.
In \autoref{sec:overview}, we give an overview of our hybrid approach. 
In~\autoref{sec:heap}, we propose a novel symbolic memory model for Rust compatible with Gillian, capable of both layout-independent reasoning about Rust memory and performing pointer arithmetic and bit-level operations. 
In \autoref{sec:mutable-borrows}, we demonstrate how to leverage Gillian's unique extensibility to encode concepts from the \emph{lifetime logic} of RustBelt and obtain a substantial degree of automation, enabling Gillian-Rust to reason about TS of mutable references. 
In \autoref{sec:functional-correctness}, we show how to embed within Gillian-Rust the ability to reason about parametric prophecies as proposed by RustHornBelt, enabling FC verification.
In \autoref{sec:anatomy}, we describe end-to-end verification of a safe-unsafe Rust program, elaborating on the interpretation of hybrid specifications into Creusot and Gillian-Rust specifications and the details of Gillian-Rust automation.
In \autoref{sec:eval}, we evaluate Gillian-Rust by verifying TS and FC of several Rust standard library types and their safe clients, demonstrating the efficiency and scalability of our hybrid approach.
Finally, we discuss the current limitations of Gillian-Rust in detail and provide a pathway towards overcoming these limitations (\autoref{sec:limitations}), place Gillian-Rust in the context of overall related work (\autoref{sec:relwork}), and give concluding remarks~(\autoref{sec:concs}).


\vspace*{-0.1cm}
\section{Overview}
\label{sec:overview}

We present our hybrid approach in more detail, show how Gillian-Rust can be used for proving a Creusot specification, and describe the structure of Gillian-Rust as an instantiation of Gillian.

\vspace*{-0.1cm}
\subsection{A hybrid approach: Creusot + Gillian-Rust}

The unmatched simplicity of Creusot specifications and the extent of its proof automation come from the fact that its proofs do not manipulate the real representation of objects, but a pure abstraction instead. Take, for example, doubly-linked lists, which are infamously difficult both to implement in Rust and to specify without separation logic~(SL). Creusot, when performing the proof for code that uses the Rust \lstinline{LinkedList} module, does not see its low-level representation but instead models the linked list as a sequence of values. This approach, made possible by the guarantees provided by safe Rust, sacrifices the ability to reason about the \lstinline{LinkedList} implementation in exchange for an efficient encoding into SMT, a high degree of automation, and no need for~SL.

\begin{figure}[!t]
    \centering
    \includegraphics[width=0.88\textwidth]{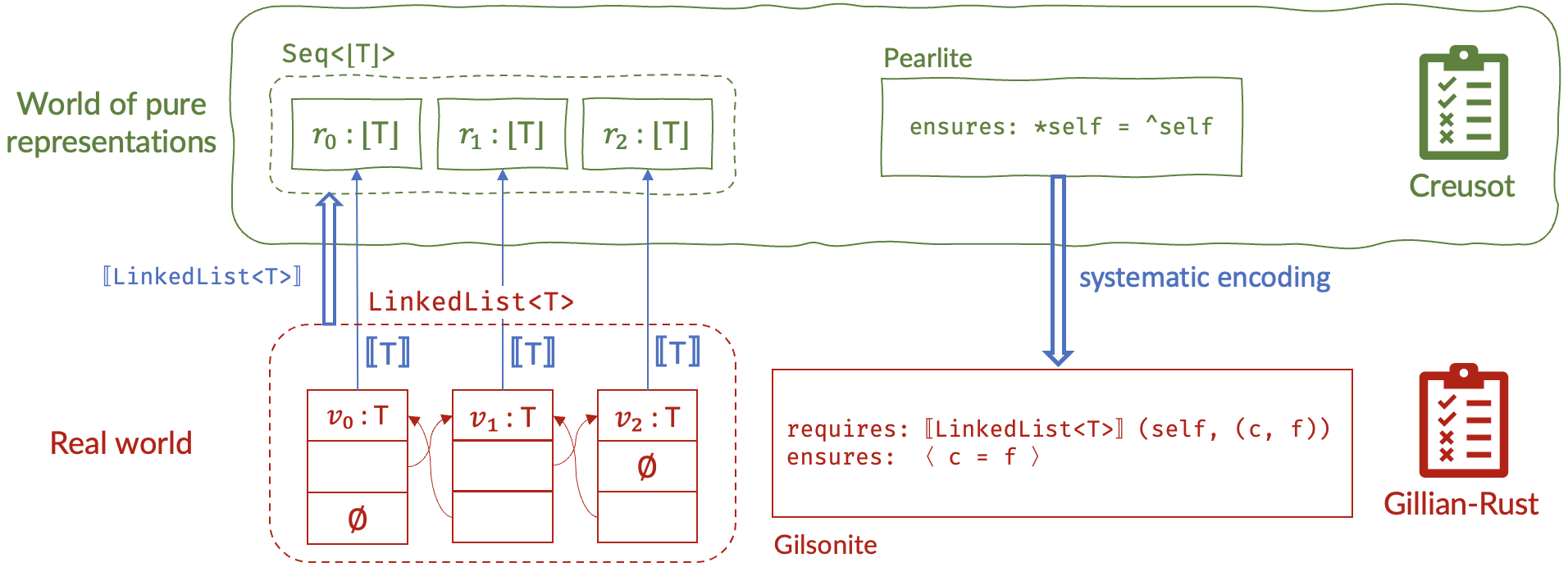}
    \vspace*{-0.2cm}
    \caption{A high-level illustration of the differences and connections between the world of pure representations, observed by Creusot, and the world of real representations, observed by RustHornBelt and Gillian-Rust.}
    \label{fig:worlds}
    \vspace*{-0.4cm}
\end{figure}

RustHornBelt provides a foundational argument for the validity of this approach by connecting the real world to Creusot's world of pure representations. This is done by providing \emph{ownership predicates}
for each type $\pty$, which describe the safety invariant that the values of this type must uphold and connect it to the associated pure representation of type $\repr{\pty}$ (cf. \autoref{fig:worlds} (left)).

To verify real-world Rust, we propose a hybrid approach where Creusot verifies all proof obligations within its reach and delegates unsafe code verification to a tool dedicated for that purpose. As such a tool does not yet exist, we develop a proof-of-concept called Gillian-Rust, which has the ability to perform SL reasoning required for the verification of unsafe code, breaking the abstraction and manipulating ownership predicates directly.

A keystone to this approach is the ability to systematically encode Creusot specification, written in an assertion language called \emph{Pearlite}, into the assertion language of Gillian-Rust, which we dub \emph{Gilsonite}, as represented in \autoref{fig:worlds} (right), and detailed in \autoref{sec:anatomy}.

\vspace*{-0.1cm}
\subsection{Example usage of Gillian-Rust}
\label{sec:overview:linkedlist}

Doubly-linked lists are notoriously difficult to implement in Rust: the presence of back edges violates the strict ownership discipline imposed by the use of mutable references. Instead, one must use mutable \emph{raw pointers}, as per the code below, making doubly-linked lists a canonical example of a data structure requiring an unsafe implementation. On top, the non-trivial invariant that the list can be integrally traversed in both directions without cycles must be upheld, as otherwise the function in charge of disposing the list would visit a node twice, thereby performing a double-free.

\begin{lstlisting}[numbers=none, basicstyle=\ttfamily\scriptsize]
struct Node<T>       { elem: T, next: Option<NonNull<Node<T>>>, prev: Option<NonNull<Node<T>>> }
struct LinkedList<T> { head: Option<NonNull<Node<T>>>, tail: Option<NonNull<Node<T>>>, len: usize }
\end{lstlisting}

We show the process of using Gillian-Rust to prove a Pearlite specification for the \rusti{push_front} function of the Rust standard library, which in-place adds an element to the front of a \linkedlist.

\begin{wrapfigure}{R}{0.46\textwidth}
\vspace*{-0.05cm}\hspace*{-0.15cm}
\begin{minipage}{0.45\textwidth}
\begin{lstlisting}[numbers=none, basicstyle=\ttfamily\scriptsize]
impl<T : Ownable> Ownable for LinkedList<T> {
  type ReprTy = Seq<T::ReprTy>;
  #[predicate] 
  fn own(self, repr: Self::ReprTy) {
    dllSeg(self.head, None, self.tail, None, repr) * 
    (self.len == repr.len()) }
}
\end{lstlisting}
\end{minipage}
\vspace*{-0.4cm}
\end{wrapfigure}

\myparagraph{Implementing Ownable}
First, we connect the real Rust structure to its pure representation used by Creusot by implementing the \lstinline{Ownable} \emph{trait}\footnote{A trait is, akin to a Haskell typeclass, a form of interface describing a list of items that can be implemented for a type.} and defining: the type of its representation, \lstinline{ReprTy} (denoted by $\repr{\cdot}$ in mathematics); and the ownership predicate, \lstinline{fn own}, which takes two parameters: the structure itself (\lstinline{self}) and the representation.
The implementation of \lstinline{Ownable} for \lstinline{LinkedList<T>} is given on the right-hand side (for the \lstinline{dllSeg} predicate, \ifbool{extendedversion}{see~\autoref{sec:heap:logic}}{see~\cite{gillian-rust-arxiv}}): its representation type is a sequence of elements of type \lstinline{T::ReprTy}. Note that, in order for this type to be properly defined, \lstinline{T} itself must implement \lstinline{Ownable}, a constraint specified using a trait bound (the \lstinline{': Ownable'} part in \lstinline{<T : Ownable>}). 

\myparagraph{Type safety (TS)}
Reasoning modularly about TS of unsafe code is challenging, and involves non-trivial implicit proof obligations. This is due to the fact that, by nature, type safety of a library is a global property, as ``unsafety'' may escape the scope of a single unsafe function~\cite{scope-unsafe-jung-2016}. Thankfully, RustBelt provides a way to reason about TS of a function in isolation.
Specifically, ownership predicates capture the invariant that must be upheld by the structure to ensure TS. In Gillian-Rust, once the ownership predicate for \lstinline|LinkedList<T>| has been defined by the user, we can verify TS of a function by simply adding the \lstinline{#[show_safety]} attribute on top, as follows:
\begin{lstlisting}[basicstyle=\ttfamily\scriptsize]
#[show_safety]
// Expands to: #[specification( requires { self.own(_) * e.own(_) } ensures { result.own(_) })]
fn push_front(&mut self, e : T) { ...implementation... }
\end{lstlisting}

This attribute expands to a Gilsonite specification which requires the ownership predicate of all input parameters to hold when entering the function, and ensures that the ownership predicate of the return value holds on function return.
Here, the function \lstinline|push_front| receives a mutable reference to a \lstinline|LinkedList<T>| as an argument, and the ownership predicate of mutable references, initially formalised in RustBelt (and later extended by RustHornBelt), is automatically derived by Gillian-Rust. The rules that allow for manipulating the ownership invariant of mutable references are challenging to automate, and we detail our approach to this in \autoref{sec:mutable-borrows}. In addition, as the function is implicitly parametrised by the lifetime of the mutable reference, a \emph{lifetime token} is added automatically by the Gillian-Rust compiled (cf.~\autoref{fig:lifetime-tokens}). Gillian-Rust is able to prove this specification fully automatically.

\myparagraph{Functional correctness} 
Next, our goal is to specify that the function actually performs the desired operation. 
This can be elegantly done in Pearlite by describing the update performed on the sequence which represents the \lstinline{LinkedList}: \textit{when the mutable reference expires}, the representation of the mutable reference will be its representation when the function is entered with the element prepended.
Representations are accessed using the postfix operator \lstinline|@|, the current value of a mutable reference is accessed using the dereference prefix operator \lstinline|*|, and the value of a mutable reference at the time it expires is accessed using the prophecy prefix operator \lstinline|^|:
\begin{lstlisting}[basicstyle=\ttfamily\scriptsize]
#[requires((*self)@.len() < usize::MAX@)] 
#[ensures((^self)@ == (*self)@.prepend(e))]
fn push_front(&mut self, e : T) { /* Implementation ... */ mutref_auto_resolve!(self) }
\end{lstlisting}

Pearlite, inspired by RustHorn~\cite{rusthorn-chcbased-verification-matsushita-2021}, uses prophecy variables and the final value operator \lstinline{^} in order to specify such a property. 
RustHornBelt provides the theory underpinning this, and we provide a high-level description of the corresponding proof techniques as well as their implementations and automation strategies in Gillian-Rust in \autoref{sec:functional-correctness}.
Using our systematic encoding, we can translate this Pearlite specification into a Gilsonite specification: this particular translation is given in \autoref{sec:anatomy}, together with further explanations.
Finally, after adding a single line which triggers a semi-automatic tactic during verification, Gillian-Rust is able to prove this specification.

\subsection{Building Gillian-Rust on top of Gillian}
\label{sec:overview:gillian}

Gillian-Rust is an instantiation of Gillian~\cite{gillian-part-multilanguage-fragososantos-2020, gillian-part-ii-maksimovic-2021}, a multi-language compositional symbolic execution platform.
To instantiate Gillian to a target language (TL), one must implement a symbolic state model of the TL in OCaml, exposing: a representation of the symbolic TL state, as  an OCaml type; \emph{actions}, which are primitive operations for manipulating the state; and \emph{core predicates}, which are the building blocks of an SL assertion language for describing states, and which allow one to write function specifications, user-defined predicates (e.g., for describing data structures), loop invariants, and proof tactics (e.g., predicate folding/unfolding). One must also implement a compiler from the TL to Gillian's intermediate language (which is a simple goto-based intermediate language parametric on the above-mentioned actions), and from TL assertions to Gillian assertions (which are parametric on the above-mentioned core predicates).

In Gillian-Rust, symbolic states have the form $\sst = (\symb{h}, \lftctx, \predctx,  \obsctx, \pcyctx)$, comprising: a symbolic heap~$\symb{h}$~(\autoref{sec:heap}); a lifetime context~$\lftctx$ (\autoref{sec:lifetimes:lftctx}); a guarded predicate context~$\predctx$ (\autoref{sec:lifetimes:guarded-preds}); an observation context~$\obsctx$ (\autoref{sec:functional-correctness:observts}); and a prophecy context~$\pcyctx$ (\autoref{sec:functional-correctness:pcyctx}).

Gillian action execution is described using judgements of the form $\acttrans{\sst}{\pc}{\mathit{act}}{\vec{\sval}}{(\sst', \sval_o)}{\pc'}$, the meaning of which is that: in the symbolic execution configuration $(\sst, \pc)$ where $\sst$ is a symbolic state and $\pc$ is a path condition (i.e. a first-order formula constraining the symbolic variables), executing action $\mathit{act}$ with arguments $\vec{\sval}$ yields a state $\sst'$, value $\sval_o$, and path condition $\pc'$. Expectedly, symbolic execution may branch, that is, executing an action may produce several outcomes.

For each core predicate $\rho$, Gillian requires two actions: the \emph{consumer}, $\consf_\rho$, which removes the resource corresponding to $\rho$ from a given symbolic state; and the \emph{producer}, $\prodf_\rho$, which does the opposite. On top, Gillian extends consumption and production to entire assertions, enabling compositionality (through reuse of function specifications) and predicate folding and unfolding. This is what makes Gillian uniquely extensible in the space of semi-automated compositional verification tools, as it allows one to automate the basic rules of their custom SL. Under the hood, consumption and production are powered by an assertion matching mechanism that enforces predictable, backtrack-free proof search~\cite{gillian-part-ii-maksimovic-2021,cse,matching-plans-frame-loow-2024}. Further, the predicate folding is almost fully automated, while predicate unfolding is performed heuristically. The strong performance of Gillian is evidenced by the verification times obtained for real-world JavaScript, C, and now Rust code.




\section{Reasoning about the real Rust heap}
\label{sec:heap}

While RustBelt provides the theoretical framework on which our work is founded, it intentionally avoids the challenge of reasoning about the real Rust heap by instead defining an operational semantics and type system for $\lambdarust$, a small lambda-calculus with a simplified memory model. For example, in $\lambdarust$, all integers are unbounded and take one cell in memory, ignoring the 12 different primitive machine integer types offered by Rust, which take between 1 and 16 bytes in memory.

The literature, from previous work on other systems programming languages such as C, already has ways of reasoning about machine integers, but Rust also comes with challenges currently undealt with. In particular, while C comes with a specific algorithm that describes and decides on the layout of structures in memory and allows for arbitrary pointer arithmetic to access structure fields, the Rust compiler provides fewer guarantees, reserving the right to re-order fields and adjust padding between them. Rust also has features that do not exist in C, such as enums (tagged unions), which offer even fewer guarantees, as Rust may manipulate fields arbitrarily to reduce the overall size of the structure without affecting expressivity, in a process called niche optimization.

So far, Rust symbolic execution tools have been working around these issues. For example,
Prusti encodes structures using the object-oriented memory model of Viper, allowing efficient field access but preventing pointer arithmetic reasoning, and 
Kani compiles Rust to a C-like representation by choosing a specific layout for each structure, dropping the guarantee that a verified program would be correct had the compiler made different layout choices authorised by the language~\cite{rust_lang_unsafe_code_guidelines_2019}.

In this section, we describe the solution provided by Gillian-Rust, which does a best-effort attempt at \emph{maintaining abstraction}---hence preserving field-access efficiency---while still allowing for pointer arithmetic by leveraging Gillian's ability to implement custom heap models directly in OCaml. We show how to encode addresses so that they are layout-independent, describe a novel representation of objects in the heap that allows for efficient automated reasoning, and present the points-to core predicate, which allows for specifying the Rust heap in Gillian-Rust.

\vspace*{-0.1cm}
\subsection{Layout-independent memory addresses}

The representation of addresses in Rust constitutes a challenge on its own. Ideally, one would prefer to reuse the one used by Gillian-C, inspired by CompCert~\cite{compcert-memory-model-leroy-2012} and also used in RustBelt, where an address is a pair $(\loc, o) \in \locs \times \nats$ of an object location (identifying a unique allocation) and an offset. However, because of the above-mentioned challenges, this representation is insufficient, as structure field access may correspond to different offsets depending on the compiler-chosen layout.

\begin{wrapfigure}{r}{0.36\textwidth}
\vspace{-0.5cm}\hspace*{-0.35cm}
\begin{minipage}{0.36\textwidth}
\[
\begin{array}{c}
\begin{array}{lll}
\loc \in \locs & e \in \widehat{\mathbb{Z}} & i, j \in \nats
\end{array}\\
\begin{array}{lll}
\proj \in \mathtt{ProjE} & ::= & +^{\pty} e \mid .^{\pty} i \mid .^{\pty.j} i \\
a \in \addrs & ::= & (l, \vec{\proj} )
\end{array}
\end{array}
\]
\end{minipage}
\vspace{-0.3cm}
\end{wrapfigure}

To overcome this issue, Gillian-Rust modifies the encoding of offsets by using sequences of \emph{projection elements} forming a \emph{projection} (we reuse the compiler's internal terminology) instead of a natural number. Specifically, a projection element represents either: an offset of $e$ times the size of the type $\pty$, where $e$ is a symbolic integer, denoted by $+^{\pty} e$; or the offset of the $i$-th field of a structure (relative w.r.t.~the beginning of the structure), denoted by $.^{\pty} i$; or the relative offset of the $i$-th field of the $j$-th variant of an enum, denoted by $.^{\pty.j} i$. 

This representation makes the interpretation of a symbolic address effectively parametric on the layout chosen by the compiler: given a layout which provides a concrete offset for each field of a structure or an enum, and a size to every type, each projection element can be interpreted as a symbolic natural number, and each projection as the sum of the interpretations of its elements.


\vspace*{-0.1cm}
\subsection{Objects in the Rust symbolic heap}
\label{sec:objects}


Our goal is to represent objects in the symbolic heap in a way that would enable us to efficiently resolve field accesses and perform only layout-independent pointer arithmetic. To this end, we propose a hybrid tree representation featuring two kinds of nodes: {\em structural nodes}, which represent a region of memory for which we know the structure but not necessarily the layout (such as Rust structures or enums), and on which no pointer arithmetic is allowed; and {\em laid-out nodes}, which are known to have an array-like layout and admit certain pointer arithmetic. For clarity of presentation, we provide a high-level description of the heap, focussing on the main functionalities and insights. 

\myparagraph{Structural nodes} Structural nodes are annotated with their type, and may be one of the~following:
\begin{itemize}[leftmargin=*]
\item a single node containing either: the special value $\mathtt{Uninit}$, representing uninitialised memory, which is illegal to read; the special value $\mathtt{Missing}$, representing memory that has been framed off; or a symbolic value;

\item a tree representing a structure, consisting of: a root (internal) node, which holds no information; and children nodes, which represent its fields; or

\item a tree representing an enum with a concrete discriminant\footnote{A symbolic enum (i.e., an enum with a symbolic discriminant) would be represented as a single node with a symbolic value.}, containing: an internal node holding said discriminant; and children nodes representing the fields of the corresponding enum variant. 
\end{itemize}

The types annotating the nodes must be \emph{sized} (i.e., must have a size known at compile-time\footnote{In contrast to unsized types, such as the slice type $[\pty]$, for which the size is only known at run-time.}), thereby providing an interpretation for each node. The \emph{load} and \emph{store} primitive operations are provided in the interface of the symbolic heap and must ensure that the validity invariants~\cite{two-kinds-invariants-jung-2018} of values written in memory are maintained (e.g., that booleans are represented by bit-patterns \lstinline{0b0} and \lstinline{0b1} only). They are also responsible for enforcing other important aspects of the Rust semantics, such as that loading a value from memory in the context of a {\em move} will deinitialise that memory.

In the diagram below, we give an example of a structure \texttt{S} and its structural node representation, comprising an internal node annotated with type \texttt{S} and two single-node children with respective values and types $(\symb{x}, {\color{tealgreen}\mathtt{u32}})$ and $(\symb{y}, {\color{tealgreen}\mathtt{u64}})$. The type of the left child, for example, indicates that it represents a region of 4 bytes in memory, and that the symbolic value $\symb{x}$ is an integer in the range $[0, 2^{32})$. We also show two potential interpretations of a structural node for \texttt{S}, depending on the compiler-chosen field ordering: the top interpretation is obtained when the ordering is from-largest-to-smallest, and the bottom when the ordering is from-smallest-to-largest, inserting the appropriate padding when needed. This structural node in particular can only be navigated using $.^{\mathtt{S}}0$ or $.^{\mathtt{S}}1$.

\begin{wrapfigure}{r}{0.31\textwidth}
  \vspace{-0.55cm}
  \begin{minipage}{0.31\textwidth}
  \begin{tabular}{c}
  \begin{lstlisting}[numbers=none]
  struct S { x: u32, y: u64 };
  \end{lstlisting}\\[-0.4cm]
  \begin{tikzpicture}[every node/.style={draw,shape=circle,minimum size=.7cm}]
  \node (S) {}
      child { node (x) { $\symb{x}$ } 
              edge from parent
              node[midway, left, draw=none] {.0} 
      }
      child { node (y) { $\symb{y}$ } 
              edge from parent
              node[midway, right, draw=none] {.1} 
      };
  \node[draw=none] (Sty) at ($(S) + (0.5, 0.4)$) {\lstinline!S!};
  \node[draw=none] (u32) at ($(x) + (0.5, 0.4)$) {\lstinline!u32!};
  \node[draw=none] (u64) at ($(y) + (0.5, 0.4)$) {\lstinline!u64!};
  \end{tikzpicture}\\[-0.1cm]
  \begin{tikzpicture}[scale=0.85]
  \node (0) at (0, .7) { 0 };
  \draw (0, 0) rectangle (2, .5) node [pos=.5] { $\symb{y}$ };
  \node (4) at (2, .7) { 8 };
  \draw (2, 0) rectangle (3, .5) node [pos=.5] { $\symb{x}$ };
  \node (6) at (3, .7) { 12 };
  \draw[pattern=north east lines, pattern color=black] (3, 0) rectangle (4, .5);
  \node (8) at (4, .7) { 16 };
  \end{tikzpicture}\\
  \begin{tikzpicture}[scale=0.85]
  \node (0) at (0, .7) { 0 };
  \draw (0, 0) rectangle (1, .5) node [pos=.5] { $\symb{x}$ };
  \node (2) at (1, .7) { 4 };
  \draw[pattern=north east lines, pattern color=black] (1, 0) rectangle (2, .5);
  \node (4) at (2, .7) { 8 };
  \draw (2, 0) rectangle (4, .5) node [pos=.5] { $\symb{y}$ };
  \node (8) at (4, .7) { 16 };
  \end{tikzpicture}
  \end{tabular}
  \end{minipage}
  \vspace*{-0.5cm}
  \end{wrapfigure}
  
\myparagraph{Laid-out nodes}
While structural nodes facilitate efficient resolution for a large majority of memory accesses, they are not a novel concept. The novelty of our approach lies in combining structural nodes with laid-out nodes, inspired by Gillian-C~\cite{gillian-part-ii-maksimovic-2021}, which describe a region of memory with an array-like layout in the sense that it allows for basic indexing pointer arithmetic.
For example, Rust arrays, which are at the core of the Rust vector type, are always laid out contiguously such that the $n$-th element of an array of type $[\pty; N]$ starts at offset $n\ *\ $\lstinline{size_of::<T>()} w.r.t~the beginning of the array, regardless of the layout of the element itself. Similarly, any integer type, say ${\color{tealgreen}\mathtt{u32}}$, can be seen as array-like as it is always represented by contiguous bytes in memory. 

\begin{figure}[!t]
\begin{minipage}{\textwidth}
\centering
\begin{tikzpicture}
  \draw[purple, dashed, rounded corners] (-.3,-.3) rectangle (3.8,1) node[pos=1] (NE) {};
  \draw (0, -.1) rectangle (1.4,0.6) node[pos=.5] { $\symb{\vec{v}}$ };
  \draw (1.4, -.1) rectangle (3.5,0.6) node[pos=.5] { \texttt{Uninit} };
  \node (0) at (0, .8) { 0 };
  \node (k) at (1.4, .8) { $\symb{k}$ };
  \node (n) at (3.5, .8) { $\symb{n}$ };
  \node[anchor=south east, color=purple] at (NE) {indexing type: \texttt{T}};
  \draw[double, thick, -{Implies[]}] (3.9,.35) -- (4.3,.35) node[midway, above] { \tiny isolate };
\end{tikzpicture}\hspace{-.1cm}%
\begin{tikzpicture}
  \draw[purple, dashed, rounded corners] (-.3,-.3) rectangle (3.8,1) node[pos=1] (NE) {};
  \draw (0, -.1) rectangle (1.4,0.6) node[pos=.5] { $\symb{\vec{v}}$ };
  \draw (1.4, -.1) rectangle (2.1,0.6) node[pos=.5] { \tiny \texttt{Uninit} };
  \draw (2.1, -.1) rectangle (3.5,0.6) node[pos=.5] { \texttt{Uninit} };
  \node (0) at (0, .8) { 0 };
  \node (k) at (1.4, .8) { $\symb{k}$ };
  \node (k+1) at (2.1, .8) { $\symb{k} + 1$ };
  \node (n) at (3.5, .8) { $\symb{n}$ };
  \node[anchor=south east, color=purple] at (NE) {indexing type: \texttt{T}};
  \draw[double, thick, -{Implies[]}] (3.9,.35) -- (4.3,.35) node[midway, above] { \tiny write };
\end{tikzpicture}\hspace{-.1cm}%
\begin{tikzpicture}
  \draw[purple, dashed, rounded corners] (-.3,-.3) rectangle (3.8,1) node[pos=1] (NE) {};
  \draw (0, -.1) rectangle (1.4,0.6) node[pos=.5] { $\symb{\vec{v}}$ };
  \draw (1.4, -.1) rectangle (2.1,0.6) node[pos=.5] { };
  \draw (2.1, -.1) rectangle (3.5,0.6) node[pos=.5] { \texttt{Uninit} };
  \node[circle, draw=black, minimum size=0.7cm, anchor=center, fill=white] (S) at (1.75, .25) { $\symb{v}'$ };
  \node (0) at (0, .8) { 0 };
  \node (k) at (1.4, .8) { $\symb{k}$ };
  \node (k+1) at (2.1, .8) { $\symb{k}+1$ };
  \node (n) at (3.5, .8) { $\symb n$ };
  \node[anchor=south east, color=purple] at (NE) {indexing type: \texttt{T}};
\end{tikzpicture}
\end{minipage}
\vspace*{-0.2cm}
\caption{Update of a laid-out node corresponding to $n\ *\ $ \lstinline{size_of::<T>()}  bytes. 
}
\label{fig:laid-out-vec}
\vspace*{-0.4cm}
\end{figure}

A laid-out node is a pair composed of a sized type (called indexing type) and a list of structural nodes each annotated with the range it occupies in multiples of the size of the indexing type. For example, \autoref{fig:laid-out-vec} (left) shows a laid-out node with indexing type \lstinline{T} and two structural nodes, the first carrying a symbolic list value $\symb{\vec{v}}$ occupying the range $[0, \symb k)$ (note that the $\symb k$ is symbolic), and the second capturing uninitialised memory occupying the range $[\symb k, \symb n)$, with $\symb k < \symb n$. 

When resolving pointer arithmetic, Gillian-Rust is able to automatically destruct and reassemble laid-out nodes, allowing for arbitrary range access and manipulation. For example, \autoref{fig:laid-out-vec} (middle) and (right) show the process of writing a single value of type \texttt{T} at the $\symb k$-th offset; this  corresponds to pushing at the end of a vector with sufficient capacity.
Gillian-Rust achieves this by first isolating the region in which the newly added value is going to be written (\autoref{fig:laid-out-vec}, middle), splitting the second node into two, and then overwriting the appropriate region (in this case, from $\symb k$ to $\symb k + 1$) with a structural node corresponding to the added value (\autoref{fig:laid-out-vec}, right), simplified for this example to be a single node.
Importantly, the indexing type does not have to match the type of each individual sub-node. For example, explicit calls to the Rust allocator API will always result in a laid-out node with indexing type \lstinline{u8} (i.e., single bytes), but can be populated with values of arbitrary other type \lstinline{T}.

\subsection{Specifying the Rust heap: the typed points-to core predicate}
\label{sec:heap:logic}

We focus on the most important core predicate used to specify heap shape with Gilsonite: the typed points-to predicate, $a \typointsto{\pty} v$, which is satisfied by a heap fragment starting from address~$a$ and containing \lstinline{size_of::<T>()} bytes, which together form a valid representation of the value $v$. The remaining core predicates are only variations on this theme and are used for specifying, for example, slices or potentially uninitialised memory.

The separation logic induced by the core predicates can be used by the verification engineer to specify a variety of predicates, pre-conditions and post-conditions. For example, the typed points-to predicate is enough to specify the ownership predicate for the \linkedlist type of the standard library, \ifbool{extendedversion}{which we now present in detail}{which is described in detail in~\cite{gillian-rust-arxiv}}.

\begin{extendedversiononly}
Using mathematical notations, the ownership predicate of the \linkedlist is defined as:

\[
\begin{array}{l@{\tbspace}l@{\tbspace}l}
\\[-0.9cm]
\ownpredlst{LinkedList<T>}(l, r) & \triangleq & \mathtt{dllSeg}\langle\pty\rangle(l\facc{head}, \mathtt{None}, l\facc{tail}, \mathtt{None}, r) * l\facc{len} = |r|\\
\mathtt{dllSeg}\langle\pty\rangle(h, n, t, p, r) & \triangleq &
     (h = n * t = p * r = []) \lor\\
 & & (\exists h', v, z, r'.~ h = \mathtt{Some}(h') * h' \typointsto{\mbox{\lstinline{Node<T>}}} \{ v, z, p \} * \ownpred{\pty}(v, r_v)\ * \\
 & & \hspace*{1.8cm} \mathtt{dllSeg}\langle\pty\rangle(z, n, t, h, r') * r = r_v::r')
\end{array}
\]

The doubly-linked-list-segment predicate, $\mathtt{dllSeg}$, is well-known from SL literature.
It receives four optional pointers, $h$, $n$, $t$, and $p$, and a sequence of values $r$.
The pointers $h$ and $t$ represent, respectively, the head and the tail pointer to the doubly-linked list, while $n$ corresponds to the $\mathtt{next}$ pointer of the tail node and $p$ to the $\mathtt{prev}$ pointer of the head node; both $p$ and $n$  equal $\mathtt{None}$ when the list segment represents the entire linked list. The sequence $r$ contains the values of the nodes in the list, ordered left-to-right.
This predicate can be reused in the context of Rust with only one adaptation: the value of each node must be owned by the list (captured by the $\ownpred{\pty}(v, r_v)$ ownership predicate), effectively making the predicate parametric invariant of the type of values that the list holds.

The segment predicate and the \rusti{LinkedList<T>} type invariant can be defined using Gilsonite as follows; note that the \rusti{->} arrows need not be annotated with the type, as type inference is performed by the Rust compiler:

\begin{lstlisting}[numbers=none]  
#[predicate]
fn dll_seg<T: Ownable>(h: Option<NonNull<Node<T>>>, n: Option<NonNull<Node<T>>>,
                       t: Option<NonNull<Node<T>>>, p: Option<NonNull<Node<T>>>,
                       r: Seq<T::ReprTy>) {
    gilsonite!(h == n * t == p * r == Seq::empty());
    gilsonite!(exists hp, z, v, rv. h == Some(hp) * hp -> Node { next: z, prev: p, element: v } * 
                                    v.own(rv) * dll_seg(z, n, t, h, r.prepend(rv)))
}

impl<T : Ownable> Ownable for LinkedList<T> {
  type ReprTy = Seq<T::ReprTy>;
  #[predicate] 
  fn own(self, repr: Self::ReprTy) -> Gilsonite {
    dllSeg(self.head, None, self.tail, None, repr) * 
    (self.len == repr.len())
  }
}
\end{lstlisting}

\end{extendedversiononly}


\section{Automating reasoning about mutable borrows}
\label{sec:mutable-borrows}

Handling mutable borrows is one of the main challenges when trying to specify and verify Rust programs in fully-safe and unsafe contexts alike.
While RustBelt~\cite{rustbelt-securing-foundations-jung-2017} provides a theoretical framework for reasoning about mutable borrows within Iris and proves its correctness in Rocq, this reasoning itself is manual and slow. In this section, we show how to leverage the unique flexibility of Gillian to automate reasoning about lifetimes and basic operations on mutable borrows.

\vspace*{-0.1cm}
\subsection{Modelling lifetimes: core predicates}
\label{sec:lifetimes:lftctx}

In Rust, a lifetime is a type-level variable representing a period of time during which a reference is valid. It is the responsibility of the {borrow checker} of the compiler to compute sound lifetimes for all references so that the ownership discipline of Rust is maintained.

In RustBelt, lifetimes are encoded as \emph{tokens} in its separation logic: the token $\alivetkn{\klft}_q$, with $0 < q \leq 1$, represents an alive lifetime $\klft$, while $\deadtkn{\klft}$ denotes that the lifetime $\klft$ has expired.
RustBelt also provides rules to reason about lifetime tokens, some of which are included below for illustrative purposes: e.g., \inferref{LftL-not-own-end} states that a lifetime cannot be alive and expired at the same time;  \inferref{LftL-end-persist} states that an expired lifetime token is persistent (i.e. it can be duplicated); while \inferref{LftL-tok-fract} states that alive lifetime tokens may be split into fractions (for $0 < q, q'$).
\begin{mathpar}
\inferrule[LftL-not-own-end]{}{\alivetkn{\klft}_q \lstar \deadtkn{\klft} \Rightarrow \lfalse}
\and%
\inferrule[LftL-end-persist]{}{\persistent{\deadtkn{\klft}}}
\and%
\inferrule[LftL-tok-fract]{}{\alivetkn{\klft}_{q + q'} \Leftrightarrow \alivetkn{\klft}_q * \alivetkn{\klft}_{q'}}
\end{mathpar}

\begin{wrapfigure}{r}{0.29\textwidth}
\vspace*{-0.6cm}
\hspace*{-0.12cm}\begin{minipage}{0.3\textwidth}
\[
\begin{array}{l@{\tbspace\in\tbspace}l@{\tbspace}l@{\tbspace}l}
\klft & \lfts & \approx & \mathcal{P}(\nats) \\
\lftctx & \lftctxs & = & \lfts \pfmap \widehat{\mathbb{R}}_{(0, 1]}^\dagger
\end{array}
\]
\vspace*{-0.5cm}
\end{minipage}
\end{wrapfigure}

A lifetime context $\lftctx$ is then a partial finite map from lifetimes to either the currently owned fraction of the lifetime token (a symbolic real number in the $(0, 1]$ interval), or an indicator of expiration, $\dagger$.

In Gillian-Rust, both kinds of tokens become core predicates, and
we demonstrate how the three RustBelt rules shown above are automated by providing an excerpt of the rules governing their consumers and producers in \autoref{fig:lifetime-tokens}.\footnote{In these rules, to avoid clutter: the judgement uses only the lifetime context instead of the entire symbolic state; and the return value is elided because both actions return unit.} While simple, these rules are illustrative of the relationship between custom consumers/producers and automation. For example, the rule \inferref{Lft-Produce-Alive-Add} adds a fraction $q$ of an alive token when a fraction $q'$ is already owned, automating the right-to-left implication of \inferref{LftL-tok-fract}. On the other hand, \inferref{Lft-Produce-Own-End} vanishes (i.e. assumes $\lfalse$) when producing an alive token in a context where the lifetime has expired, automating \inferref{LftL-not-own-end}. Similarly, in the consumer/producer paradigm, a core predicate is made persistent when its producer is idempotent and its consumer does not modify memory. Hence, together, rules \inferref{Lft-Consume-Exp} and \inferref{Lft-Produce-Exp-Dup} automate \inferref{LftL-end-persist}.

\begin{figure}[!t]
\begin{mathpar}
\inferrule[Lft-Produce-Alive-Add]{
\lftctx(\klft') = q'\\
\pc \vdash (\klft = \klft' \land 0 < q \land q + q' \leq 1)\\
\lftctx' = \lftctx\mapupdate{\klft}{q + q'}}
{
	\prodtrans{\lftctx}{\pc}{\alivelftcp}{\klft,q}{\lftctx'}{\pc}
}
\and%
\inferrule[Lft-Produce-Own-End]{\lftctx(\klft') = \dagger \\ \pc \vdash (\klft = \klft')}
{
	\prodtransvanish{\lftctx}{\pc}{\alivelftcp}{\klft,q}
}%
\and%
\inferrule[Lft-Consume-Exp]{
    \lftctx(\klft') = \dagger \\
    \pc \vdash (\klft = \klft')
}
{
	\constrans{\lftctx}{\pc}{\deadlftcp}{\klft}{\lftctx}{\pc}
}%
\and%
\inferrule[Lft-Produce-Exp-Dup]{
    \lftctx(\klft') = \dagger \\
    \pc \vdash (\klft = \klft')
}{
\prodtrans{\lftctx}{\pc}{\deadlftcp}{\klft}{\lftctx}{\pc}
}%
\end{mathpar}
\vspace*{-0.3cm}
\caption{Consumer and producer rules for lifetime tokens (simplified, excerpt)}
\label{fig:lifetime-tokens}
\vspace*{-0.2cm}
\end{figure}

\subsection{Modelling full borrows: guarded predicates}
\label{sec:lifetimes:guarded-preds}

In Rust, a mutable reference of a value of type $\pty$ during lifetime $\klft$, denoted by $\mutref{\klft} \pty$, corresponds to \emph{temporary} ownership of the reference and the value it points to. To model such a behaviour, RustBelt introduced full borrows, denoted by $\fbw{\klft} P$, which are higher-order predicates denoting that the resource described by assertion $P$ is borrowed during lifetime $\klft$. In RustBelt, where ownership predicates do not expose a pure representation, the ownership predicate of a mutable reference $p$ and the key rules for manipulating mutable borrows are as follows:
\begin{mathpar}
\ownpred{\mutref{\klft}{\pty}}(p) \triangleq \fbw{\klft}( \exists v.~ p \mapsto v \lstar \ownpred{\pty}(v) )%
\and
\inferrule[LftL-borrow-acc]{}{
	\fbw{\klft} P \lstar \alivetkn{\klft} \irupdate \later P \lstar ( \later P \irupdate \fbw{\klft} P \lstar \alivetkn{\klft})
}
\end{mathpar}
In particular, \inferref{LftL-borrow-acc}
states that one may \emph{open a borrow} by temporarily giving up the corresponding lifetime token, and may later \emph{close that borrow} after having reformed the invariant, at which point the token is recovered. Crucially, having to reform the invariant inside a borrow is what ensures that a callee function which is given a borrow may not cause undefined behaviour in the future, and every borrow must eventually be closed, as the lifetime token is required at the time it expires. In Gillian-Rust, the view shift operator present in the \inferref{LftL-borrow-acc} rule is realised via guarded predicate unfolding, introduced shortly, whereas the later modality, $\later$, is omitted; in \autoref{sec:limitations}, we provide a justification for the soundness of this approach.

Full borrows raise two main challenges for a semi-automated tool such as Gillian: \textbf{1)} it needs to reason about higher-order predicates; and \textbf{2)} it needs to automatically understand when to open and close borrows in common proof patterns. We now present the two key insights behind the encoding and automation of reasoning about full borrows in Gillian-Rust.

\myparagraph{Compiling higher-orderness away}
While program proofs do make use of higher-order rules such as \inferref{LftL-borrow-acc}, they only use them with a specific, finite set of instantiations. For example, when proving \lstinline{pop_front_node}, one only needs to manipulate the particular borrow predicate
 corresponding to the ownership predicate 
 $\refmutownpredlst{\klft}{LinkedList<T>}$. When using the Gilsonite API, a user may instantiate the full borrow assertion using the \lstinline{#[borrow]} attribute. For instance, the ownership predicate for mutable references is defined as follows in the Gilsonite library:

\begin{lstlisting}[numbers=none, basicstyle=\ttfamily\scriptsize]
impl<T> Ownable for &mut T { #[borrow] fn own(self) -> Gilsonite { exists v. (self -> v) * v.own() } }
\end{lstlisting}

\noindent
obtaining an ownership predicate for mutable references of type \lstinline{T}. Note that such predicates can be defined parametrically, using a generic type; when required for a more specific type, such as \lstinline{LinkedList<T>}, they will be instantiated at compilation time.

Finally, ownership predicates for type parameters are compiled to abstract predicates, that is, predicates that cannot be unfolded, a well-known trick in the world of semi-automated tools. This ensures that if a specification has been proven using a type parameter $\pty$, then this type parameter can be instantiated with any other type to obtain a new trusted specification, with the instantiation happening at the call site that requires it.

\myparagraph{Leveraging known automations for borrow access} The key insight to automating borrow access is the understanding that borrows behave very similarly to standard predicates encoded in a semi-automated SL-based verification tool. In particular, VeriFast, Viper, and Gillian all support predicates of the form $(\upred, \vec{\sval}) \in (\strs \times \mathsf{List}({\svals}))$, where each predicate consists of a name $\upred$ (normally a string) and parameters $\vec{\sval}$. Predicates of this form are said to be {\em folded} and each of the above-mentioned tools maintains a list of predicates as part of their state.


Each of these tools also comes with two ghost commands that allow users to manipulate folded predicates: $\mathtt{unfold}$ and $\mathtt{fold}$. In particular, $\mathtt{unfold}$ removes a predicate stored in its folded form from the state and produces its definition in its place, whereas $\mathtt{fold}$ is its dual, consuming the predicate's definition from the state and adding its folded form to the state.

One may notice the similarity between the borrow access rule and the folding and unfolding of predicates: when closed, both borrows and folded predicates act as abstract tokens that can be exchanged for the resource they contain. The only distinction is the ``cost'' of unfolding: none for predicates, and a lifetime token for borrows.

A guarded predicate context $\predctx \in \mathsf{List}(\strs \times {\color{purple}\lfts} \times \mathsf{List}(\svals))$ is a list of predicates which are annotated with a lifetime such that its token is the cost for their opening. It exposes two actions: $\mathtt{gunfold}$/$\mathtt{gfold}$, which respectively behave like $\mathtt{unfold}$/$\mathtt{fold}$ apart from the fact that they \linebreak consume/produce that guarding lifetime token, and produce/consume an additional opaque \emph{closing token}, denoted by $\closetkn{\upred}(\klft, q, \vec{x})$, which embodies the closing update $( P \irupdate \fbw{\klft} P \lstar \alivetkn{\klft}_q)$.

\begin{wrapfigure}{r}{0.44\textwidth}
\vspace*{-0.5cm}
\begin{minipage}{0.44\textwidth}
\begin{mathpar}
\inferrule[Unfold-Guarded]{
  p\facc{predDefs}[\upred(\klft, \vec{x})] = P\\
  {\color{purple} \constrans{\sst}{\pc}{\alivelftcp}{\alft, q}{\sst'}{\pc'}}\\
  \sst' = (\smem', \predctx') \quad \upred(\alft, \vec{\sval}) \in \predctx'\\
  \predctx'' = \predctx' \setminus \upred(\alft, \vec{\sval}) \quad \sst'' = (\smem', \predctx'')\\
  {\color{purple} P' = P * \closetkn{\upred}(\klft, q, \vec{v})}\\
  \acttrans{\sst''}{\pc'}{\prodf}{P'[\vec{x}/\vec{\sval}]}{\sst'''}{\pc''}
}{
  p \vdash \acttrans{\sst}{\pc}{\mathtt{gunfold}}{\upred(\alft, \vec{\sval})}{\sst'''}{\pc''}
}
\end{mathpar}
\end{minipage}
\vspace{-0.3cm}
\end{wrapfigure}

The \inferref{Unfold-Guarded} rule describes successful execution of $\mathtt{gunfold}$. For clarity, we decompose symbolic states into a pair $(\mu, \predctx)$, where $\mu$ represents the remaining components. In addition, we write {\color{purple} in purple} elements of the rule which are novel with respect to the more classic $\mathtt{unfold}$ rule. Finally, this command is performed in the context of a program $p$, where $p\facc{predDefs}$ maps predicates to their definitions.

This encoding of full borrows has one important advantage: Gillian comes with years of experience in automating separation logic proofs, including heuristics that are able to decide when to automatically unfold or fold predicates as required by the analysis. By encoding borrows in the above way, we can immediately leverage those heuristics and allow for automatic opening and closing of full borrows.
In particular, proving the type safety of \lstinline{LinkedList::pop_front} and \lstinline{LinkedList::push_front} becomes completely automatic once the safety invariants of \lstinline{LinkedList} has been properly specified as in \autoref{sec:heap:logic}.

\subsection{Proving safety of borrow extraction}
\label{sec:borrow-extract}

Unfortunately, opening and closing are not the only operations that one needs when working with full borrows. We identify several recurring patterns in unsafe Rust programs and provide ways of instantiating lemmas that allow us to analyse code that uses these patterns.

In particular, borrow extraction---the process of cutting a borrow up into a smaller borrow---is a common pattern in unsafe Rust programming, and every data-structure module of the standard library provides at least one function that uses this pattern (e.g., \lstinline{LinkedList::front_mut} or \lstinline{Vec::get_mut}). In fact, borrow extraction is the most idiomatic way of modifying an element of a~collection.
Most often, implementing such a function is unsafe, as incorrect borrow extraction could break the safety guarantees of Rust. For example, consider the case in which the \lstinline{LinkedList} library implementer creates a \lstinline{first_node_mut} function, which returns a mutable reference not to the first element (\lstinline{&mut T}), but to the first node (\lstinline{&mut Node<T>}), which contains the first element as well as \lstinline{next} and \lstinline{prev} pointers (\autoref{fig:mutrefs}, left). Then, \textbf{using only safe code}, a client function could modify the \lstinline{next} pointer to point to the node itself, creating a cycle in the list. As explained in \autoref{sec:overview:linkedlist}, this would certainly lead to an undefined behaviour, although not during the execution of \lstinline{first_node_mut} itself.

\begin{wrapfigure}{r}{0.28\textwidth}
\vspace*{-0.4cm}
\begin{minipage}{0.28\textwidth}
\centering
\begin{tikzpicture}[scale=0.8]
  \matrix[every node/.style={draw,rectangle,minimum width=1cm,minimum height=6mm}] (a) {
    \node (n-1){$v$};\\
    \node (next1){next}; \\
    \node (prev1){prev};\\
  };
  \node [anchor=south,color=red] (mutNode) at ($(n-1.north east) + (0.4, 0.4)$) {\footnotesize $\mutref{}\mathtt{Node}\langle\pty\rangle$};
  \draw[color=red, ultra thick] (n-1.north west) rectangle (prev1.south east);
  \draw[->, ultra thick,color=red] (mutNode.south) -- (n-1.north east);
  
  \matrix[every node/.style={draw,rectangle,minimum width=1cm,minimum height=6mm}] (c) at (2, 0) {
    \node (n+1){$v$};\\
    \node (next3){next};\\
    \node (prev3) {prev}; \\
  };
  \node [anchor=south,color=lightgreen] (mutT) at ($(n+1.north east) + (0.4, 0.4)$) {\footnotesize $\mutref{}\pty$};
  \draw[color=lightgreen, ultra thick] (n+1.north west) rectangle (n+1.south east);
  \draw[->, ultra thick,color=lightgreen] (mutT.south) -- (n+1.north east);
\end{tikzpicture}
\vspace{-0.2cm}
\caption{An invalid and a valid \lstinline{LinkedList} mutable reference.}
\label{fig:mutrefs}
\vspace{-0.2cm}
\end{minipage}
\end{wrapfigure}

On the other hand, returning a mutable reference to the first element (\lstinline{&mut T}), as per \autoref{fig:mutrefs} (right), is not an issue, with the intuition being that one can \emph{remove} the resource associated with the element and obtain a \emph{remainder}. To that remainder one can then add any other element that satisfies the invariant of $\pty$, recovering a structure satisfying the \lstinline{LinkedList} invariant.  This principle is embodied by the \inferref{borrow-extract} rule (which we have proven in Iris using RustBelt), where $P$ is the invariant of the \lstinline{LinkedList}, $Q$ is the invariant of $\pty$, and $Q \wand P$ is the remainder. In addition, the rule allows one to add a persistent context if it is required for performing the extraction. For example, in the case of the \lstinline{LinkedList}, the extraction of the first node is only possible if it is not empty (i.e. if the \lstinline{head} pointer is not \lstinline{None}, which would be captured in that persistent context).

\begin{wrapfigure}{r}{0.29\textwidth}
\vspace*{-0.55cm}
\hspace*{-0.2cm}
\begin{minipage}{0.29\textwidth}
\small
\begin{mathpar}
\inferrule[borrow-extract]{
    \persistent{F}\\
    F * P \Rightarrow Q * (Q \wand P)
}{
    F * \alivetkn{\klft}_q * \fbw{\klft}P \irupdate \fbw{\klft}Q  * \alivetkn{\klft}_q
}
\end{mathpar}
\end{minipage}
\vspace{-0.6cm}
\end{wrapfigure}

Using the Gilsonite API, users may instantiate the ghost command that performs the view shift in the conclusion of the \inferref{borrow-extract} rule by specifying the borrow predicates $\fbw{\klft}P$ and $\fbw{\klft}Q$ as well as the persistent assertion $F$, as illustratively done below\footnote{The \lstinline{list_ref_mut_frozen} predicate is a borrow predicate obtained from the ownership predicate of \lstinline{&mut LinkedList} by freezing existentials corresponding to the \lstinline{head}, \lstinline{tail} and \lstinline{len} fields of the structure. Freezing existentials is a common strategy for extracting borrows, supported by the Gilsonite API. \ifbool{extendedversion}{To avoid cluttering the main presentation, we present it in~\autoref{sec:appendix-freeze}}{For lack of space, we present this in~\cite{gillian-rust-arxiv}}.}:

\begin{lstlisting}[numbers=none, basicstyle=\ttfamily\scriptsize]
#[extract_lemma( forall head, tail, len, p. assuming { head == Some(p) } // ~{\color{gray} $F$}~
  from { list_ref_mut_frozen(list, head, tail, len) } //  ~{\color{gray} $\fbw{\kappa} P$ }~
  extract { Ownable::own(&mut (*p.as_ptr()).element) } //  ~{\color{gray} $\fbw{\kappa} Q$ }~
)]
fn extract_head<T: Ownable>(list: &mut LinkedList<T>); // Implicitly parametric on ~{\color{gray} $\klft$}~
\end{lstlisting}

Gillian itself cannot prove that \inferref{borrow-extract} holds or manipulate borrows using such a rule. Instead, the Gillian-Rust compiler produces two lemmas: one corresponding to the rule conclusion, which is marked as trusted and left unproven, and one corresponding to the rule hypotheses, which needs to be proven. As the rule itself has been proven to hold in Iris, the Gillian-Rust meta-theory therefore ensures that if we prove the second lemma, the first lemma also has to hold. 

To automatically prove this second kind of lemmas, we have extended Gillian with the ability to reason about magic wands, adapting the related work on Viper~\cite{sound-automation-magic-dardinier-2022}, to Gillian's parametric separation logic; the details of this extension are out of scope of this presentation.


\section{Functional correctness and prophetic reasoning}
\label{sec:functional-correctness}

While the ability to manipulate full borrows is enough to verify type safety of programs that make use of mutable references, it is not enough to prove functional correctness of these programs. 
In particular, the rule \inferref{LftL-borrow-acc} presented previously enforces that the \textbf{same} invariant be used to close the full borrow, effectively losing the information that the value was updated.

Specifying functional correctness of programs manipulating mutable references is, in itself, a challenge, as it requires the ability to specify properties which shall only hold \emph{in the future}, that is, at the time when the borrow expires.
Thankfully, this challenge has been addressed by previous work: Prusti~\cite{leveraging-rust-types-astrauskas-2019} introduced pledges and RustHorn~\cite{rusthorn-chcbased-verification-matsushita-2021} introduced prophecy variables, later used in Creusot.
However, only the latter has been given a foundational formalisation in RustHornBelt~\cite{rusthornbelt-semantic-foundation-matsushita-2022}, an extension of RustBelt which describes how prophetic specifications interact with full borrows.

We next recall the workings of RustHornBelt and show how its concepts are encoded in Gillian-Rust.
To conclude our technical presentation, we show how Pearlite specifications are compiled to Gilsonite, explaining how unsafe proof goals can be delegated by Creusot to Gillian-Rust.

\subsection{Representations, parametric prophecies, and observations}
\label{sec:functional-correctness:rusthornbelt}

In order to reason about functional correctness within the framework of RustBelt, RustHornBelt extends ownership predicates with an additional  parameter corresponding to a pure mathematical \emph{representation} of the value. Given a type $\pty$, the type of its representation is denoted by $\repr{\pty}$. For example, a value of type \lstinline{LinkedList<T>} is represented by a sequence of which each element is the representation of the element at the corresponding index in the list, i.e. $\repr{\mbox{\lstinline{LinkedList<T>}}} = \mbox{\lstinline{Seq<}}\repr{\mbox{\lstinline{T}}}\mbox{\lstinline{>}}$.

Mutable references, on the other hand, are represented as a pair of representations of the inner type  (i.e., $\repr{\mbox{\lstinline{&mut T}}} = \repr{\pty} \times \repr{\pty}$), where the first element denotes the value to which the mutable reference currently points, and the second denotes the value it will have at the time it expires.

\begin{wrapfigure}{r}{0.47\textwidth}
\vspace{-0.45cm}
\begin{minipage}{0.47\textwidth}
\[
\begin{array}{l}
\ownpred{\mutref{\klft}{\pty}}(p, r) \triangleq \exists x\text{ s.t. } r.\!\!^{\star}2 = \uparrow\! x.~\vobs{x}(r.\!\!^{\star}1)\ * \\
\qquad \fbw{\klft}(\exists v, a.~ p \mapsto v * \ownpred{\pty}(v, a) * \pctrl{x}(a))
\end{array}
\]
\end{minipage}%
\vspace{-0.4cm}%
\end{wrapfigure}

RustHornBelt then proposes an ownership predicate for mutable references which exposes this representation, using a notion of \emph{parametric prophecies}.
A prophecy variable $x$ is attached to the mutable reference, and the second element of the representation pair $r$ is the future value of this prophecy, denoted by $\uparrow\!x$.

\begin{wrapfigure}{r}{0.28\textwidth}
\vspace{-0.54cm}
\begin{minipage}{0.28\textwidth}
\begin{mathpar}
\inferrule[Mut-Agree]{}{
\vobs{x}(a) * \pctrl{x}(a') \vdash a = a'
}%
\and%
\inferrule[Mut-Update]{}{
\vobs{x}(a) * \pctrl{x}(a) \Rrightarrow \\ \quad \vobs{x}(a') * \pctrl{x}(a')
}
\end{mathpar}
\end{minipage}%
\vspace{-0.4cm}%
\end{wrapfigure}

In addition, there are two connected resources respectively called \emph{value observer}, denoted by $\vobs{x}$, and \emph{prophecy controller}, denoted by $\pctrl{x}$, which together provide a solution to the problem of information loss when closing a full borrow.
In particular, the observer maintains the last-observed current value and, when the borrow opens, the previously-lost value of the representation $a$ is recovered through the \inferref{Mut-Agree} rule. Before closing a borrow again, the verification engineer may use the \inferref{Mut-Update} rule to update the value of the prophecy variable to match the new representation. 

Lastly, RustHornBelt introduces \emph{observations}, denoted by $\observt{\psi}$, where $\psi$ is a pure assertion containing information known about prophecy values.
Observations act as a second layer of truth, preventing future information from leaking into the separation logic and creating~paradoxes.

\vspace*{-0.1cm}
\subsection{Key idea: parametric prophecies and symbolic execution}
\label{sec:functional-correctness:observts}

In order to encode prophecies into Iris, RustHornBelt wraps the entire execution into a reader monad. In simple terms, execution is performed within a context which preemptively captures an assignment for the future value of each existing prophecy variable (i.e., a map $\pcyvars \rightarrow \mathit{Value}$).

One of the key ideas presented in this work comes from noticing that symbolic execution in the Gillian meta-theory can be formalised using an environment of the same nature, of type $\mathit{SVar} \rightarrow \mathit{Value}$, which assigns a concrete interpretation to each symbolic variable.
Therefore, parametric prophecies appear to be closer to symbolic variables than they are to prophecy variables formalised by Jung et al.~\cite{future-ours-prophecy-jung-2019}.
This intuition suggests that one may use the same process to reason about prophecy variables as for symbolic variables, and ideally fit them into the same framework.
In symbolic execution, each state carries a \emph{path condition} $\pc$, a pure formula which accumulates all currently-known constraints about the existing symbolic variables, while for prophecy variables, it is the observations that play this role of constraint accumulator.
The core idea behind encoding prophecy variables follows from this remark: observations can simply take the shape of a secondary path condition, implemented as a custom resource algebra in OCaml within the Gillian framework, making calls to the Gillian solver when required.

To this end, we introduce a new custom resource algebra in Gillian which consists of only one symbolic expression, called \emph{observation context} and denoted by $\obsctx \in \mathit{Obs}$. The observation context may depend on both prophecy variables and symbolic variables. \autoref{fig:obs-cons-prod} (top) presents some of the rules that apply to observations in RustHornBelt, while \autoref{fig:obs-cons-prod} (bottom) shows Gillian-Rust consumer and producer rules for the successful cases. Again, for clarity of presentation, we elide the non-needed components of the state and the return values.

\begin{figure}
\begin{mathpar}  
\inferrule[Obs-merge]{}{
    \observt{\psi} \lstar \observt{\psi'} \vdash \observt{\psi \land \psi'}
}
\and%
\inferrule[Proph-Sat]{}{
\observt{\psi} \Rightarrow \exists \prophasn.\ \prophasn(\psi)
}
\and%
\inferrule[Proph-True]{}{(\forall \prophasn.\ \prophasn(\psi)) \Rightarrow \observt{\psi}}
\\
\inferrule[Observation-Produce]{
  \pc \land \obsctx \land \obsctx' \text{ SAT}
}
{
    \prodtrans{\obsctx}{\pc}{\observt{\cdot}}{\obsctx'}{\obsctx \land \obsctx'}{\pc}
}%
\and%
\inferrule[Observation-Consume]{
  (\pc \land \obsctx \Rightarrow \obsctx') \text{ VALID}
}
{
	\constrans{\obsctx}{\pc}{\observt{\cdot}}{\obsctx'}{{\obsctx}}{\pc}
}%
\end{mathpar}
\vspace*{-0.3cm}
\caption{Excerpts: observation rules from RustHornBelt (top) and observation consumer/producer rules (bottom)}
\label{fig:obs-cons-prod}
\vspace*{-0.2cm}
\end{figure}

\inferref{Obs-merge} indicates that our model of observations as a single symbolic expression is appropriate, and that framing on a new observation amounts to simply conjuncting it with the current observation.
In addition, \inferref{Proph-Sat} tells us that if an observation holds, then at least one prophecy assignment must satisfy it.
Together, these rules instruct us how to implement the producer for observations: if the conjunction of the path condition, current observation, and new observation is satisfiable, then we can add the produced observation to our current one (cf. \inferref{Observation-Produce}).
Finally, \inferref{Proph-True} states that anything that is true independently of prophecy variables can be captured as an observation, that is, anything that is true outside of the prophetic world is also true within it.
With our approach, this means that the path condition can be used seamlessly as part of our observations when needed, embodied in the \inferref{Observation-Consume} rule: when checking if an observation $\obsctx'$ holds, we check that it is entailed by the current path condition and observation.

\subsection{Value observers and prophecy controllers}
\label{sec:functional-correctness:pcyctx}

Value observers and prophecy controllers provide yet another opportunity to leverage the flexibility of Gillian and implement a custom resource algebra.
In particular, we entirely automate the \inferref{Mut-Agree} rule by defining a \emph{prophecy context} $\pcyctx \in \pcyvars \rightarrow \mathit{Expr} \times \bools \times \bools$ as a map that associates each prophecy variable with its current value and two Booleans, which correspond to the ownership of the value observer and of the prophecy controller in the state.

Below, we provide rules for successfully producing a value observer into the state; the production rules for the prophecy controller are analogous and therefore elided:
\begin{mathpar}
\inferrule[VObs-Produce-Without-Controller]{
  x \notin \dom(\pcyctx)\quad
  \pcyctx' = \pcyctx\mapupdate{x}{(a, \top, \bot)}
}{
  \prodtrans{\pcyctx}{\pc}{\vobs{}}{x,a}{\pcyctx'}{\pc}
}%
\and%
\inferrule[VObs-Produce-With-Controller]{
  \pcyctx(x) = (a', \bot, \top)\quad
  \pcyctx' = \pcyctx\mapupdate{x}{(a', \top, \top)}
}{
  \prodtrans{\pcyctx}{\pc}{\vobs{}}{x,a}{\pcyctx'}{\pc \land (a = a')}  
}%
\end{mathpar}
In particular, producing $\vobs{x}(a)$ in a prophecy context which does not already contain any binding for the prophecy variable $x$ will bind $x$ to the triple $(a, \top, \bot)$, thereby encoding that the current value for the prophecy is $a$, that its value observer is in the context, but not its prophecy controller.
On the other hand, if the controller with value $a'$ already exists in the current state, that is, if the prophecy context already has the triple $(a', \bot, \top)$ bound to $x$, then the Boolean flag corresponding to the presence of the corresponding value observer is set to true without modifying the current value and we learn that $a = a'$, in the form of an additional constraint added to the path condition.

However, this does not automate the \inferref{Mut-Update} rule: after modifying the contents of a mutable reference \lstinline{p: &mut T}, one still needs to apply this rule to be able to close the mutable borrow.
The current Gillian implementation does not allow full automation of this process, but we are able to provide the \inferref{Mut-Auto-Update} lemma, which the verification engineer can apply by writing \lstinline{p.prophecy_auto_update()}, and which updates the current value of the prophecy by automatically choosing the appropriate value that will allow the borrow to be closed.

\begin{wrapfigure}{r}{0.36\textwidth}
\vspace{-0.3cm}
\begin{minipage}{0.36\textwidth}
\inferrule[Mut-Auto-Update]{}{
\ownpred{\pty}(v, a') * \vobs{x}(a) * \pctrl{x}(a) \Rrightarrow \\\\
\qquad \ownpred{\pty}(v, a') * \vobs{x}(a') * \pctrl{x}(a')
}
\end{minipage}
\\[2mm]
\begin{minipage}{0.36\textwidth}
\inferrule[MutRef-Resolve]{}{
\ownpred{\mutref{\klft}\pty}(p, (a, a')) \irupdate \observt{a = a'}
}
\end{minipage}
\vspace{-0.35cm}
\end{wrapfigure}

Finally, Gillian-Rust also provides a manual way of \emph{resolving} mutable 
references, as described by \inferref{MutRef-Resolve}, which, as proposed by RustHornBelt, allows us to obtain an observation of the equality between the current value of the prophecy and its future value at the time where the corresponding mutable reference expires.

\myparagraph{Borrow extraction with prophecies}
When manipulating the ownership predicate of a mutable reference with prophecies in the style of RustHornBelt, 
the rule for extracting sub-borrows must be adapted to perform \emph{partial resolution} of the prophecy. The corresponding rule is substantially more complex than \inferref{borrow-extract}, but it yields the same level of automation and we have proven it correct in the Rocq development of RustHornBelt. \ifbool{extendedversion}{To avoid clutter, we present it in \autoref{sec:appendix-extract}}{For lack of space, we present it in~\cite{gillian-rust-arxiv}}.


\section{Anatomy of a hybrid proof : Merge Sort}
\label{sec:anatomy}

In this section, we present a detailed example of a hybrid proof, showing how we can use Creusot and Gillian-Rust to prove the correctness of a Merge Sort implementation that uses doubly-linked lists.
We briefly cover the safe implementation and its verification in Creusot, and then explain how we interface with Gillian-Rust to prove correctness of associated unsafe operations.

\begin{figure}[!t]
\begin{lstlisting}[numbers=left, numbersep=5pt, xleftmargin=2em, basicstyle=\ttfamily\scriptsize]
#[pearlite::ensures(sorted((^l)@) && l@.permutation_of((^l)@))] ~\label{line:merge-sort:perm}~
pub fn merge_sort(l: &mut LinkedList<i32>) { /* Standard impl. using split and merge */ }

#[pearlite::ensures(inp@.permutation_of(result.0@.concat(result.1@)))] ~\label{line:split:perm}~
fn split(inp: &mut LinkedList<i32>) -> (LinkedList<i32>, LinkedList<i32>) {
  let old_inp = snapshot!(inp);
  let mut (left, right, push_left) = (LinkedList::new(), LinkedList::new(), true);
  let mut popped = snapshot! { Seq::EMPTY };
  #[pearlite::invariant(
    popped.concat(inp@).ext_eq(old_inp@) && popped.permutation_of(left@.concat(right@))
  )]
  while let Some(i) = inp.pop_front() {
      popped = snapshot! { popped.push(i) };
      snapshot!({perm_right::<i32>; perm_left::<i32>});
      if push_left { left.push_front(i); } else { right.push_front(i); };  
      push_left = !push_left;
  }
  (left, right)
}

#[pearlite::requires(sorted(l@))] #[pearlite::requires(sorted(r@))]
#[pearlite::ensures(sorted(result@) && result@.permutation_of(l@.concat(r@)))]
fn merge(l: &mut LinkedList<i32>, r: &mut LinkedList<i32>) -> LinkedList<i32> { ... }
\end{lstlisting}
\vspace*{-0.6cm}
\caption{A fragment of our Merge Sort algorithm, implemented using doubly-linked lists}
\label{fig:merge-sort}
\vspace*{-0.3cm}
\end{figure}

\myparagraph{Writing a hybrid proof}
Following the approach outlined in \S\ref{sec:overview}, we divide the work as follows:
(1) Creusot is responsible for verifying the safe parts (here, the Merge Sort algorithm itself), which normally constitute the great majority of the code; while
(2)~Gillian-Rust is responsible for verifying the unsafe parts (here, the doubly-linked list operations), which are normally more low-level and perform more complex but smaller operations such as manipulation of pointers or uninitialised memory.
In Figure \ref{fig:merge-sort}, we present a fragment of our Merge Sort implementation.
For space reasons, we elide the (standard) implementations of \lstinline{merge_sort} and \lstinline{merge}, focusing instead on the \lstinline{split} function, which takes a mutable borrow to a linked list and splits it into two halves.

In Creusot, unsafe types such as \rusti{LinkedList<T>} are treated as \emph{opaque types}, on which no operations can be performed. To reason about them, Creusot axiomatises their representation function using a \rusti{ShallowModel} trait, and the Pearlite\footnote{Pearlite is a first-order logic, including the standard connectives as well as support for functions and predicate definitions.} specifications of their APIs are assumed as axioms. 
We can access this shallow model through its associated operator \lstinline{@}. Using this model operation, we specify the postcondition of the \lstinline{split} function as per line \ref{line:split:perm} of Figure \ref{fig:merge-sort}, stating that the concatenation of the two resulting lists is a permutation of the input list. Operations on mutable borrows are specified using the \emph{final} operator \rusti{^}, which accesses the prophecy of a mutable reference.
In line~\ref{line:merge-sort:perm}, we specify that the initial value (\lstinline{(*l)@}) of the list is a permutation of its final value (\lstinline{(^l)@}).

  \begin{figure}[!t]
  \begin{lstlisting}[basicstyle=\ttfamily\scriptsize]
  pub struct LinkedList<T> { ... }
  
  impl<T : Ownable> LinkedList<T> {
    #[hybrid::ensures(forall<x : _> result == Some(x) ==> Seq::singleton(x).concat((^self)@) == (*self)@)]
    #[hybrid::ensures(result == None ==> ^self == *self && self@.len() == 0)]
    pub fn pop_front(&mut self) -> Option<T> { ... }
  
    #[hybrid::requires(self@.len() < usize::MAX@)]
    #[hybrid::ensures(Seq::singleton(e).concat((*self)@) == (^self)@)]
    pub fn push_front(&mut self, e: T) { ... }
  
    #[hybrid::ensures((*self)@.push(e) == (^self)@)]
    pub fn push_back(&mut self, e: T) { ... }
  }
  \end{lstlisting}
  \vspace*{-0.8cm}
  \caption{The \lstinline{LinkedList} library used by our Merge Sort algorithm}
  \label{fig:linked-list}
  \vspace*{-0.2cm}
  \end{figure}

In Figure \ref{fig:linked-list}, we present the specification of the \lstinline{LinkedList} library used by our Merge Sort.
We use the \lstinline{hybrid::requires} and \lstinline{hybrid::ensures} attributes to specify, respectively, the pre- and post-conditions of the \lstinline{pop_front}, \lstinline{push_front}, and \lstinline{push_back} functions.
These attributes act as the bridge between Pearlite and Gilsonite, in that from them, using the compilation mechanism presented shortly, we are able to generate the Gilsonite specification expected by Gillian-Rust. For example, for \rusti{push_front}, we will end up with the following specifications:
\begin{lstlisting}[basicstyle=\ttfamily\scriptsize]
  // Pearlite specification 
  #[pearlite::requires(self@.len() < usize::MAX@)]
  #[pearlite::ensures(Seq::singleton(e).concat((*self)@) == (^self)@)]
  // Gilsonite specification   
  #[gilsonite::specification(forall s_repr, e_repr.
    requires { self.own(s_repr) * e.own(e_repr) $ s_repr.0.len() < Int::from(usize::MAX) $ }
    exists r_repr. ensures { ret.own(r_repr) * $Seq::singleton(e_repr).concat(s_repr.0) == s_repr.1$ }
  )]
  pub fn push_front(&mut self, e: T) { ... }
\end{lstlisting}

Verification of the complete Merge Sort and accompanying Linked List implementation is performed by successively running \lstinline{cargo creusot} and \lstinline{cargo gillian} to generate the proof obligations for Creusot and Gillian-Rust, respectively, which are then discharged by running the appropriate backends: Why3 for Creusot and the Gillian-Rust backend for Gillian-Rust.

\myparagraph{Compilation of Creusot specifications}
To compile Creusot specifications to Gilsonite, we first need to interpret Creusot's types in Gillian-Rust.
Recall that we interpret Rust types using their \emph{representations}, and that \rusti{LinkedList<T>} is interpreted via the \rusti{Ownable} trait in Gillian-Rust as \rusti{gillian_rust::Seq<T::ReprTy>}.
In addition, we must interpret the \emph{logical} types of Creusot, which is also done by defining appropriate instances of \rusti{Ownable}: in particular, the \rusti{creusot::Seq<T>} type of Creusot, just like \rusti{LinkedList<T>} or Rust, is interpreted as \rusti{gillian_rust::Seq<T::ReprTy>}.
Like Creusot and RustHornBelt, we interpret mutable borrows as a pair of the representation of the value and a prophecised value, so that \rusti{&mut LinkedList<T>} is interpreted as \rusti{(Seq<T::ReprTy>, Seq<T::ReprTy>)}.

\begin{wrapfigure}{R}{0.44\textwidth}
\vspace*{-0.1cm}
\begin{minipage}{0.44\textwidth}
\[\hspace*{-0.3cm}\begin{array}{c}
\pearlitet{
  P
}{
  ~\fnkw\ \mathtt{f}\langle \klft \rangle(x_1: \pty_1, \dots, x_n: \pty_n) \rightarrow \pty_{\ret}~
}{
  Q
} \\
\implies \\
\mlinegilsonitet{
  \left(\bigoast_{i=1}^{n} \ownpred{\pty_i}(x_i, m_i)\right)
  \lstar
  \observt{P[x_i/m_i]}
  \lstar
  \alivetkn{\klft}_q
}
{
  \fnkw\ \mathtt{f}\langle \klft \rangle(x_1: \pty_1, \dots, x_n: \pty_n) \rightarrow \pty_{\ret}
}
{\begin{array}{c}
\exists m_\ret.\
  \ownpred{\pty_{\ret}}(\ret, m_{\ret})~\lstar \\
  \observt{Q[x_i/m_i][\ret/m_{\ret}]}
  \lstar 
  \alivetkn{\klft}_q
  \end{array}
}
\end{array}\]
\vspace*{-0.4cm}
\end{minipage}
\end{wrapfigure}
Specification interpretation is done by \emph{elaboration}, the general schema of which is given on the right.
We require ownership of every function argument, associating each with a representation value, and in the end, we own the result, again associated with a representation value.
We then place the preconditions and postconditions into prophecy observations, substituting occurrences of Rust variables with their corresponding representation values.
Following this process, we obtain the Gilsonite specification for \rusti{pop_front} given earlier.

\myparagraph{Gillian-Rust in action: \rusti{LinkedList::push_front}}
To complete our tour of hybrid verification, we explain how Gillian-Rust leverages its features presented in the previous sections to prove the Pearlite specification of \rusti{push_front} method of \linkedlist. In Figure~\ref{fig:impl-push-front}, we give the full implementation of \rusti{push_front}, together with the auxiliary \rusti{push_front_node} method. We provide a specification only for the former, as Gillian-Rust can simply symbolically execute the~latter.

\begin{figure}
\begin{lstlisting}[numbers=left, numbersep=5pt, xleftmargin=2em, basicstyle=\ttfamily\scriptsize]
#[gilsonite::specification( ... )]
pub fn push_front(&mut self, elt: T) {
    self.push_front_node(Box::new(Node::new(elt))); ~\label{line:push-front:call}~
    mutref_auto_resolve!(self); // <- Single additional annotation required ~\label{line:unique}~
}

fn push_front_node(&mut self, mut node: Box<Node<T>>) { unsafe {
    node.next = self.head; node.prev = None;~\label{line:push-front-node:next}~
    let node = Some(Box::leak(node).into()); ~\label{line:push-front-node:leak}~
    match self.head { ~\label{line:push-front-node:match}~None => self.tail = node, Some(head) => (*head.as_ptr()).prev = node, } ~\label{line:push-front-node:match-end}~
    self.head = node;
    self.len += 1;  ~\label{line:push-front-node:len}~
} }
\end{lstlisting}
\vspace{-0.5cm}
\caption{Implementation of \lstinline{push_front}}
\label{fig:impl-push-front}
\vspace{-0.3cm}
\end{figure}

When execution starts, the state contains: \textbf{a)} the ownership predicate for a mutable reference to a \linkedlist at reference \rusti{self}, with representation \rusti{self_repr}; \textbf{b)} the ownership predicate for the element \rusti{elt} of type \rusti{T}; \textbf{c)} an observation that the length of the representation of the linked list is less than \rusti{usize::MAX}; and \textbf{d)} a lifetime token corresponding to the lifetime of the mutable reference \rusti{self}.

First, in line~\ref{line:push-front:call}, the function allocates a new owned pointer, \rusti{Box}, which contains a new node constructed from the element \rusti{elt}, with previous and next pointers set to \rusti{None}. This pointer is immediately passed to the auxiliary function \rusti{push_front_node}.

In line~\ref{line:push-front-node:next}, the access to \rusti{self.head} requires ownership of the corresponding location in memory, which is currently hidden in the full borrow contained in the resource \textbf{a)}. Thanks to the encoding of full borrows presented in \autoref{sec:lifetimes:guarded-preds}, Gillian-Rust automatically \emph{opens the borrow} by applying the \inferref{Unfold-Guarded} rule, losing ownership of the lifetime token (resource \textbf{d)}), but obtaining ownership of the value contained at address \rusti{self} as well as the entire linked list, together with the prophecy controller corresponding to its representation.

The following three lines perform in-place heap updates, all handled automatically by Gillian-Rust, as per \autoref{sec:heap}. Note that the matching of the value of \rusti{self.head} in line~\ref{line:push-front-node:match} and its dereferencing to access its \rusti{prev} field requires unfolding the \rusti{dllSeg} predicate once, also done automatically.

Next, in line~\ref{line:push-front-node:len}, the \rusti{len} field of the list is updated, potentially resulting in an overflow. The current path condition is not sufficient to prove its absence and execution branches into a correct path where the overflow does not happen and an incorrect path that implicitly calls a panic. Before panicking, Gillian-Rust always checks that the current path condition (here, the overflow condition) does not contradict the observation, using the \inferref{Proph-Sat} rule (which entails that $\observt{\lfalse} \Rightarrow \lfalse$). Here, the observation, our resource \textbf{c)}, contradicts the overflow, and the incorrect path is discarded.

Next, \rusti{push_front_node} returns, and the \rusti{mutref_auto_resolve!} annotation on line~\ref{line:unique} tells Gillian-Rust to apply the \inferref{Mut-Update} and \inferref{MutRef-Resolve} rules in sequence. The former requires the invariant of the linked list to have been restored, with a new representation. At this point, Gillian-Rust automatically folds the \rusti{dllSeg} predicate twice, once to revert the unfolding previously performed, and once to push the newly-added node and its ownership predicate (resource \textbf{b)}) to its front. Then, Gillian-Rust folds the ownership predicate of the linked-list, checking that the first and final pointer are \rusti{None}, and that its length field corresponds to the length of its new representation, which is \rusti{self_repr} with \rusti{elt_repr} prepended to it. \inferref{Mut-Update} is then successfully applied, updating the prophecy controller and observer to match the new representation.

When applying \inferref{MutRef-Resolve}, Gillian-Rust understands that the borrow needs to be closed. Since the invariant of the linked list has been correctly restored, the full borrow is automatically closed, and the lifetime token is recovered. \inferref{MutRef-Resolve} then discards the resource corresponding to the mutable reference (including the full borrow), and produces the observation required to prove the postcondition of the function.

Finally, the obtained state is matched against the postcondition, which requires: \textbf{1)} ownership of the return value, which is vacuously owned as the return type is unit; \textbf{2)} the lifetime token that was recovered when closing the borrow; and \textbf{3)} the observation obtained by applying \inferref{Mut-Update}. As the postcondition is satisfied, the specification is verified, and can be soundly used in Creusot.


\section{Evaluation}
\label{sec:eval}

\begin{wrapfigure}{r}{0.41\textwidth}
  \vspace*{-0.3cm}
  \centering
  \begin{tabular}{@{}l@{\hspace*{0.1cm}}>{\ttfamily}r@{\hspace*{0.25cm}}>{\ttfamily}r@{\hspace*{0.25cm}}>{\ttfamily}r@{\hspace*{0.25cm}}>{\ttfamily}r@{}}
  \toprule
  & \textrm{VP} & \textrm{eLoC} & \textrm{aLoC} & \textrm{Time} \\
  \midrule
  EvenInt & TS/FC & 47 & 13 & 0.04s \\
  LP      & TS    & 32 & 40 & 0.03s  \\
  LP      & FC    & 43 & 56 & 0.04s  \\
  LinkedList  & TS  & 130 & 176 & 0.24s \\
  LinkedList  & FC  & 130 & 227 & 0.45s  \\
  MiniVec  & FC    & 140 & 59 & 1.35s  \\  
  Vec   & TS      & 294 & 44 & 1.08s  \\
  Vec   & FC      & 294 & 107  & 2.57s \\
  \midrule
  \bottomrule
  \end{tabular}
  \vspace*{-0.3cm}
\end{wrapfigure}

We used our hybrid verification pipeline to perform a number of case studies, all making use of internally unsafe modules (IUMs); the results are shown in the table to the right. For each analysed IUM, we give: the number of executable lines of code (eLoc) and lines of annotations (specifications/predicate definitions/lemmas/proof tactics, aLoc); the type of properties verified (VP), with functional correctness (FC) subsuming type safety (TS); and the verification time. We note that verifying only TS allows for the use of a simpler encoding, which eschews prophecies to track value information. To our knowledge, this is the the first verification of TS and FC of unsafe code from the Rust standard library---a subset of the \linkedlist and \rusti{Vec} modules (with caveats for the latter)---with no or minor modifications to the original source code. All experiments were performed single-threaded, on a MacBook Pro 2019, with 16GB Memory and a 2.3GHz 9-Core Intel Core i9 processor. 

\myparagraph{EvenInt}
We start from a case study provided as part of the RefinedRust~\cite{refined-rust} evaluation. \rusti{EvenInt} is a structure that only contains a single value of type \rusti{i32}, and its ownership invariant requires the value to be even. We copy all applicable functions from this case study, eliding those that make use of shared references (cf.~\autoref{sec:limitations}), and verify them in Gillian-Rust by giving Creusot specifications that correspond to the RefinedRust specifications provided. These functions include 2~unsafe functions (one constructor and one mutator) and 3~safe functions (two constructors and one mutator). We verify Creusot specifications for the three safe functions, and purposefully do not write specifications for the unsafe functions as they are not required by Gillian-Rust, reducing the annotation overhead. Full details about the verified functions can be found \ifbool{extendedversion}{in~\autoref{sec:appendix-evenint}}{in~\cite{gillian-rust-arxiv}}.

The total verification time of Gillian-Rust for the \rusti{EvenInt} study is \textbf{0.04s}, several orders of magnitude faster than the \textbf{4m36s} of RefinedRust. To hint at the level of automation, verifying the safe mutator requires a single line of annotation with Gillian-Rust to resolve the prophecy (cf. line~\ref{line:unique} of Figure~\ref{fig:impl-push-front}). In contrast, RefinedRust requires of the user to manually write a Rocq proof that if $i$ is an even integer, then $i + 1 + 1$ is still a even integer.

\myparagraph{LinkedPair} Next, we verified TS and FC of a ``linked-pair'' data-structure (LP) that we developed as a tutorial example for Gillian-Rust, the details of which we omit given space constraints. 

\myparagraph{LinkedList}
Next, we verified TS and FC of a subset of the \linkedlist API from the Rust standard library, extracted from commit \texttt{ad2b34d0} (04/12/23) of the official Rust repository. The only modifications made were to add annotations required for verification as well as to manually inline calls to \rusti{Option::map}, whose parameter is a closure, which are not yet supported by the Gillian-Rust compiler. 
Once these are added, there will be no need for additional annotations, as Gillian-Rust will be able to symbolically execute them like any other function, without requiring a specification.

Using the ownership predicate given in \autoref{sec:overview:linkedlist} and the \texttt{dllSeg} predicate given in~\ifbool{extendedversion}{\autoref{sec:heap:logic}}{\cite{gillian-rust-arxiv}}, we prove FC of six functions: \rusti{new}, \rusti{push_front}, \rusti{pop_front}, \rusti{push_back}, \rusti{pop_back}, and \rusti{front_mut}. 
The total verification time is \textbf{0.72s}, including verification of auxiliary proofs generated by the \rusti{extract_lemma} macro, as well as two additional lemmas required for proving \rusti{push_back} and \rusti{pop_back}. 
These lemmas, in particular: change the traversal direction of \texttt{dllSeg} (from head-to-tail to tail-to-head, and vice-versa); are not Rust-specific, but rather essential primitives for any doubly-linked-list formalisation in SL; and are written in Rust and proven within Gillian-Rust without requiring the use of external tools. 

\myparagraph{MiniVec} Next, we verified a subset of the API for \rusti{MiniVec}, a simple implementation of the \rusti{Vec} module used by RefinedRust as a case study. Using specifications that provide similar guarantees to those proven by RefinedRust, we verify FC of the \rusti{new}, \rusti{with_capacity}, \rusti{push}, \rusti{pop}, \rusti{get_mut}, and \rusti{get_unchecked_mut} functions, as well as a simple associated client function. Our hybrid pipeline performs verification in \textbf{1.35s}, 1.28s for Gillian-Rust and 0.07s for Creusot, in contrast with the \textbf{30m40s} of RefinedRust.

\myparagraph{Vec}
Next, we verify the \rusti{Vec} implementation from the Rust standard library (same commit as for \linkedlist), targeting the same functions as for \rusti{MiniVec}. In addition, we also verify \rusti{index_mut}, which performs a similar operation to \rusti{get_unchecked_mut}, but adds a safety check and performs access in memory through slice indexing instead of raw pointer arithmetics. Verifying both of these functions ensures that we correctly support these two different ways of accessing memory in Rust.


The source code of \rusti{Vec} module is substantially more complex than that of \rusti{MiniVec}, explaining why the verification of this module takes longer than the verification of \rusti{MiniVec}, yielding (in our opinion, a still reasonable) \textbf{1.08s} for TS and \textbf{2.57s} for FC.

It is important to note that \rusti{Vec} performs \emph{untyped} allocations by explicitly providing the size of the allocation in bytes, yielding a raw pointer to an uninitialised array of bytes. The pointer is then cast to a pointer to the vector element type and is used to store the typed values. 
In this process, the corresponding Gillian-Rust heap object has some nodes indexed using the \rusti{u8} type and other nodes indexed using the vector element type (cf.~\autoref{sec:objects}), showcasing its resilience to low-level operations.

\myparagraph{Vec-related Caveats}
Our verification of \rusti{Vec} and \rusti{MiniVec} comes with three caveats. First, we disallow zero-sized types (ZSTs) as types of vector elements. Both \rusti{Vec} and \rusti{MiniVec} are special-cased for ZSTs, in which case there is no allocation and the vector is simply a counter for the length. However, Gillian-Rust is not able to express the ownership invariant for ZSTs, as Gillian is untyped and cannot exhibit \emph{the only representative of the ZST type}. We will overcome this limitation by allowing the Gillian-Rust state model to produce these representatives. This does mean, however, that RefinedRust considers several more execution paths for MiniVec than Gillian-Rust.

Next, the borrow extraction lemma for FC of \rusti{get_index_mut} and \rusti{index_mut} requires the proof of a magic wand that Gillian-Rust cannot yet automate, and is left unproven for now. We will add support for manually specifying extract-lemmas proofs when the tool is unable to automate them.

Finally, we slightly modified the source code of the standard library \rusti{Vec} module, in the following ways. First, the vector type is parametric on an allocator, which we remove and perform all allocation by calling the Gillian-Rust allocator. We also inline calls to functions such as \rusti{Result::map}, which receive a closure as parameter, due to the above-mentioned lack of support for closures. Finally, the real implementation of \rusti{index_mut} when using \rusti{usize} as an index is hidden behind a few layers of trait indirection; we manually inline layers so that \rusti{index_mut} is a single function.

\myparagraph{Hybrid Verification}
We argue that a \emph{hybrid} approach combining Gillian-Rust with Creusot enables higher performance and flexibility in verification. 
To validate this, it is essential to answer two questions: (1) ``Can Gillian-Rust effectively verify Creusot-style specifications?''; and (2) ``Can those specifications then be efficiently used from Creusot's perspective?''. 

In \autoref{sec:anatomy}, we presented our \rusti{hybrid} macros, which act as a bridge between the two tools, interpreting the specifications appropriately as either Pearlite or Gilsonite.
We used these macros to specify and verify the the examples presented above, conclusively answering (1). 
The code generated by the hybrid macros is often identical to the raw Gilsonite specification we would write by hand. 
We have noticed no impact on verification times caused by use of the hybrid macros. 

Our answer to (2)~comes in two parts. Firstly, we note that Creusot provides the \rusti{creusot_contracts} crate, which provides standard, trusted specifications for common Rust types through its \rusti{extern_spec!} macro.
These specifications are either identical or semantically equivalent to the ones proved by Gillian-Rust and at most a safe wrapper would be required by Creusot to prove the entailment.

Secondly, we implemented and verified several safe programs using specifications obtained by Gillian-Rust and observed favourable verification times: \textbf{Merge Sort} (55 lines of specification, 56 lines of generic lemmas about permutations missing from Creusot's standard library, and 68 executable lines of code), taking 6.3s (wall) 28.7s (user) to verify; \textbf{Gnome Sort} (6 lines of specification and 17 executable lines of code), taking 2.6s (wall), 4.6s (user) to verify; and \textbf{Right Pad} (11 lines of specification and 12 executable lines of code), taking 0.6s (wall) 0.4s (user) to verify.


\section{Limitations and Future work}
\label{sec:limitations}

Gillian-Rust is still a proof-of-concept, demonstrating, together with Creusot, the \emph{viability} of hybrid Rust verification. We outline our current limitations below, noting that most of the improvements that we believe are required for removing the `proof-of-concept' label are purely engineering challenges, but all together requiring substantial time and manpower.

\myparagraph{Unimplemented features} The Gillian-Rust compiler does not support all constructs of the MIR AST, as it was implemented by need. One such construct, for example, are closures, for which the extension should be straightforward, as Gillian-Rust already supports dynamic function calls.

Additionally, Gillian-Rust can reason about at most one lifetime \emph{in specifications} (multiple lifetimes are already supported in function bodies).
This does not allow us to verify, for instance, iterators such as \lstinline|IterMut<'_>|. The solution lies in extracting annotated lifetimes from the Rust compiler, and may involve creating a custom borrow-checker pass or modifying the Rust compiler itself.

\myparagraph{Meta-theory simplifications} The meta-theory of Gillian-Rust builds on that of RustBelt and RustHornBelt, with two simplifications that are beyond the scope of this project and that do not limit what Gillian-Rust can do, but rather leave small gaps in the justification of its~soundness. 

First, we remove `later' modalities due to Gillian's non-step-indexed separation logic, believing that if Gillian were to be formalised in Iris, our ghost commands would `take a step', justifying our approach. Second, RustHornBelt type definitions ensure the absence of causal loops in prophecy variables, by requiring an additional proof obligation that is not required by Gillian-Rust. However, Gillian enforces an ins-to-outs data flow of predicate parameters~\cite{matching-plans-frame-loow-2024} which, we believe, naturally enforces this constraint. However, formalising these properties within RustHornBelt would require a deep embedding of the assertion language, which would enable the formalisation of the dataflow analysis performed by Gillian.

\myparagraph{Unexplored topics}
We have not yet addressed shared references and their ownership predicates, which would require defining a \rusti{Shareable} trait and adding support for fractured borrows and non-atomic borrows to Gillian-Rust. To support fractured borrows, the Gillian-Rust heap would need to be extended with fractional permissions, which was done in the past for other state models without apparent roadblocks. We would also need to implement the behaviour of opening shared borrows, which is doable by following the blueprint we give for full borrows. Non-atomic should be simpler to implement, as they are more similar to full borrows, with an additional token in the guard.

In addition, while our specifications apply in concurrent contexts, we do not address concurrency-specific constructs or thread-safe types (e.g., \rusti{Send}/\rusti{Sync} proof obligations).

Finally, we do not model StackedBorrows\cite{stacked-borrows-aliasing-jung-2019} or TreeBorrows\cite{tree-borrows-villani-2025}, noting that no current theoretical framework integrates these models with the semantic typing of RustBelt.


\section{Related work}
\label{sec:relwork}

As our focus is on reasoning about unsafe Rust, we provide an overview of, to our knowledge, the only other four tools capable of performing such reasoning, none of which explores the idea of hybrid verification. Given this focus, we do not address in detail the many tools other than Creusot for verifying safe Rust (e.g., Prusti~\cite{leveraging-rust-types-astrauskas-2019}, Aeneas~\cite{aeneas-rust-verification-ho-2022}, or Flux~\cite{flux-liquid-types-lehmann-2022}). 

\myparagraph{RefinedRust} In line with RefinedC~\cite{refinedc-automating-foundational-sammler-2021}, RefinedRust~\cite{refined-rust} allows users to annotate functions with refinement types and interactively verify functional correctness (FC) of Rust programs with unsafe code. It compiles real-world Rust into an intermediate representation shallowly embedded in Rocq. Its trusted computing base is smaller than that of Gillian-Rust and it reuses RustBelt's lifetime logic to perform \emph{foundational} proofs, extending it with new techniques for automating and simplifying reasoning about unsafe code, in some instances automating reasoning that requires annotations in Gillian-Rust. Some of these automations may be worth incorporating into our work.

However, while RefinedRust can verify FC of unsafe code, it does not explore hybrid verification, the key feature of our approach, meaning that its verification of safe code will be substantially slower and less automatic.
Moreover, our preliminary evaluation suggests that Gillian-Rust is several orders of magnitude faster than RefinedRust, even for unsafe verification.

\myparagraph{VeriFast for Rust}
Rahimi Foroushaani and Jacobs~\cite{modular-formal-verification-foroushaani-2022} describe a Rust front-end for VeriFast~\cite{quick-tour-verifast-jacobs-2010} which provides a way of verifying semantic type safety for unsafe Rust. This frontend covers an extensive set of unsafe Rust features, and it has been used to verify type safety for a large corpus of real unsafe Rust code (in comparison to other unsafe Rust verification tools, including Gillian-Rust). For instance, it has been used to prove numerous proof obligations for the linked list module of the standard library.
However, this frontend focuses solely on verifying type safety of unsafe code and does not explore hybrid verification or new automations (e.g., for opening and closing borrows).

\myparagraph{Verus} In contrast to Gillian-Rust, VeriFast, and RefinedRust, Verus~\cite{verus-verifying-rust-lattuada-2023} does not use separation logic (SL) but rather linear ghost types to encode ownership properties.
This approach allows it to leverage the borrow checker of the Rust compiler to drastically improve the encoding into SMT. 
It also means that writing proofs \emph{feels like} writing Rust code, providing a familiar user experience. In short, Verus is great for verifying code that is written as target for verification (sometimes called proof-oriented programming).

However, Verus does not support traditional raw pointers, meaning that it is not able to verify `traditional' unsafe code.
For example, using Verus, one cannot verify the standard library implementation of \lstinline{LinkedList} in the way that we propose.
Instead, Verus developers have verified their own implementation of the \lstinline{LinkedList} library, keeping track of linear ghost objects (denoted by \lstinline{PointsTo<T>}, a Verus primitive) to implement links between nodes.
In that sense, Verus could be considered a verifier shallowly embedded in an extension of Rust, rather than a Rust verifier.

In addition, Verus does not yet support reasoning about functions that return mutable references, which we support in Gillian-Rust thanks to our RustHornBelt-inspired SL foundations.

\myparagraph{Kani}
Kani~\cite{how-open-source--2023} is an industrial-strength bounded model checker for Rust, which compiles an impressively large fragment of Rust to the intermediate representation ingested by CBMC~\cite{tool-checking-ansic-clarke-2004} for its analysis. However, it does not propose solutions to the challenges solved by our work: Kani cannot verify type safety; it picks a specific layout for each structure; and it treats all safe and unsafe code in the same way, not leveraging the safe Rust guarantees to enhance analysis.

\section{Conclusions}
\label{sec:concs}
We have introduced a hybrid approach to end-to-end verification of real-world Rust programs, in which, through a separation of concerns, the safe and unsafe parts of the code are handled by two different tools, each specialised for their task at hand. We have demonstrated the feasibility of this approach by connecting Creusot, a state-of-the-art automatic verification tool for safe Rust, with Gillian-Rust, a novel proof-of-concept semi-automatic verification tool for unsafe Rust. As part of the design and implementation of Gillian-Rust, we have shown how the complex concepts underpinning reasoning about unsafe Rust, such as lifetime logic and prophetic reasoning, can be brought from the interactive world of RustBelt and RustHornBelt to the world of compositional symbolic execution. We have conducted case studies that have demonstrated that Gillian-Rust is able to verify functional correctness of real-world unsafe Rust, including (to our knowledge, for the first time) code extracted from the Rust standard library, with high automation and in times several orders of magnitude faster than existing work. We have also shown that  specifications verified by Gillian-Rust can be re-used by Creusot, demonstrating the feasibility of our hybrid approach.

\section*{Data-Availability Statement}

The supplementary material, including our implementation of Gillian-Rust and the various case studies described in \autoref{sec:eval}, is available online~\cite{artifact-hybrid-approach-ayoun-2025}.

\begin{acks}
We would like to thank the reviewers, as well as Bart Jacobs and John Wickerson, whose comments have improved the overall quality of the paper. 
We would also like to thank the Rust Verification Workshop community for their welcoming environment and insightful discussions over the years.
Ayoun was supported by the Amazon Research Award `Gillian-Rust: Unbounded Verification for Unsafe Rust Code'.
Gardner was supported by the EPSRC Fellowship `VetSpec: Verified Trustworthy Software Specification' (EP/R034567/1).
\end{acks}

\bibliographystyle{ACM-Reference-Format}
\bibliography{gillian-rust}


\begin{thebibliography}{34}


\ifx \showCODEN    \undefined \def \showCODEN     #1{\unskip}     \fi
\ifx \showDOI      \undefined \def \showDOI       #1{#1}\fi
\ifx \showISBNx    \undefined \def \showISBNx     #1{\unskip}     \fi
\ifx \showISBNxiii \undefined \def \showISBNxiii  #1{\unskip}     \fi
\ifx \showISSN     \undefined \def \showISSN      #1{\unskip}     \fi
\ifx \showLCCN     \undefined \def \showLCCN      #1{\unskip}     \fi
\ifx \shownote     \undefined \def \shownote      #1{#1}          \fi
\ifx \showarticletitle \undefined \def \showarticletitle #1{#1}   \fi
\ifx \showURL      \undefined \def \showURL       {\relax}        \fi
\providecommand\bibfield[2]{#2}
\providecommand\bibinfo[2]{#2}
\providecommand\natexlab[1]{#1}
\providecommand\showeprint[2][]{arXiv:#2}

\bibitem[Astrauskas et~al\mbox{.}(2020)]%
        {how-programmers-use-astrauskas-2020}
\bibfield{author}{\bibinfo{person}{V. Astrauskas}, \bibinfo{person}{C. Matheja}, \bibinfo{person}{F. Poli}, \bibinfo{person}{P. Müller}, {and} \bibinfo{person}{A.~J. Summers}.} \bibinfo{year}{2020}\natexlab{}.
\newblock \showarticletitle{{How do Programmers use Unsafe Rust?}}
\newblock \bibinfo{journal}{\emph{Proceedings of the ACM on Programming Languages}} \bibinfo{volume}{4}, \bibinfo{number}{OOPSLA} (\bibinfo{year}{2020}), \bibinfo{pages}{136:1--136:27}.
\newblock


\bibitem[Astrauskas et~al\mbox{.}(2019)]%
        {leveraging-rust-types-astrauskas-2019}
\bibfield{author}{\bibinfo{person}{V. Astrauskas}, \bibinfo{person}{P. Müller}, \bibinfo{person}{F. Poli}, {and} \bibinfo{person}{A.~J. Summers}.} \bibinfo{year}{2019}\natexlab{}.
\newblock \showarticletitle{{Leveraging Rust Types for Modular Specification and Verification}}.
\newblock \bibinfo{journal}{\emph{Proceedings of the ACM on Programming Languages}} \bibinfo{volume}{3}, \bibinfo{number}{OOPSLA} (\bibinfo{year}{2019}), \bibinfo{pages}{147:1--147:30}.
\newblock


\bibitem[Ayoun et~al\mbox{.}(2025)]%
        {artifact-hybrid-approach-ayoun-2025}
\bibfield{author}{\bibinfo{person}{Sacha-{\'E}lie Ayoun}, \bibinfo{person}{Xavier Denis}, \bibinfo{person}{Petar Maksimovi{\'c}}, {and} \bibinfo{person}{Philippa Gardner}.} \bibinfo{year}{2025}\natexlab{}.
\newblock \bibinfo{title}{Artifact: {{A}} Hybrid Approach to Semi-Automated {{Rust}} Verification}.
\newblock \bibinfo{howpublished}{Zenodo}.
\newblock
\urldef\tempurl%
\url{https://doi.org/10.5281/zenodo.15183201}
\showDOI{\tempurl}


\bibitem[Clarke et~al\mbox{.}(2004)]%
        {tool-checking-ansic-clarke-2004}
\bibfield{author}{\bibinfo{person}{E. Clarke}, \bibinfo{person}{D. Kroening}, {and} \bibinfo{person}{F. Lerda}.} \bibinfo{year}{2004}\natexlab{}.
\newblock \showarticletitle{A {Tool} for {Checking} {ANSI}-{C} {Programs}}. In \bibinfo{booktitle}{\emph{TACAS'04}}. \bibinfo{publisher}{Springer}, \bibinfo{pages}{168--176}.
\newblock


\bibitem[Dardinier et~al\mbox{.}(2022)]%
        {sound-automation-magic-dardinier-2022}
\bibfield{author}{\bibinfo{person}{T. Dardinier}, \bibinfo{person}{G. Parthasarathy}, \bibinfo{person}{N. Weeks}, \bibinfo{person}{P. Müller}, {and} \bibinfo{person}{A.~J. Summers}.} \bibinfo{year}{2022}\natexlab{}.
\newblock \showarticletitle{Sound {Automation} of {Magic} {Wands}}. In \bibinfo{booktitle}{\emph{CAV'22}}. \bibinfo{publisher}{Springer International Publishing}, \bibinfo{pages}{130--151}.
\newblock


\bibitem[Denis and Jourdan(2023)]%
        {specifying-verifying-higherorder-denis-2023}
\bibfield{author}{\bibinfo{person}{X. Denis} {and} \bibinfo{person}{J.-H. Jourdan}.} \bibinfo{year}{2023}\natexlab{}.
\newblock \showarticletitle{Specifying and {Verifying} {Higher}-order {Rust} {Iterators}}. In \bibinfo{booktitle}{\emph{TACAS'23}} \emph{(\bibinfo{series}{Lecture {Notes} in {Computer} {Science}})}. \bibinfo{pages}{93--110}.
\newblock


\bibitem[Denis et~al\mbox{.}(2022)]%
        {creusot-foundry-deductive-denis-2022}
\bibfield{author}{\bibinfo{person}{X. Denis}, \bibinfo{person}{J.-H. Jourdan}, {and} \bibinfo{person}{C. March{\'{e}}}.} \bibinfo{year}{2022}\natexlab{}.
\newblock \showarticletitle{{Creusot: A Foundry for the Deductive Verification of Rust Programs}}. In \bibinfo{booktitle}{\emph{FMSE'22}}. \bibinfo{pages}{90--105}.
\newblock


\bibitem[Fragoso~Santos et~al\mbox{.}(2020)]%
        {gillian-part-multilanguage-fragososantos-2020}
\bibfield{author}{\bibinfo{person}{J. Fragoso~Santos}, \bibinfo{person}{P. Maksimović}, \bibinfo{person}{S.-É. Ayoun}, {and} \bibinfo{person}{P. Gardner}.} \bibinfo{year}{2020}\natexlab{}.
\newblock \showarticletitle{{Gillian, Part I: A Multi-Language Platform for Symbolic Execution}}. In \bibinfo{booktitle}{\emph{PLDI'20}}. \bibinfo{pages}{927--942}.
\newblock
\showISBNx{978-1-4503-7613-6}


\bibitem[G\"{a}her et~al\mbox{.}(2024)]%
        {refined-rust}
\bibfield{author}{\bibinfo{person}{L. G\"{a}her}, \bibinfo{person}{M. Sammler}, \bibinfo{person}{R. Jung}, \bibinfo{person}{R. Krebbers}, {and} \bibinfo{person}{D. Dreyer}.} \bibinfo{year}{2024}\natexlab{}.
\newblock \showarticletitle{{RefinedRust: A Type System for High-Assurance Verification of Rust Programs}}.
\newblock \bibinfo{journal}{\emph{Proceedings of the ACM on Programming Languages}} \bibinfo{volume}{8}, \bibinfo{number}{PLDI}, Article \bibinfo{articleno}{192} (\bibinfo{year}{2024}), \bibinfo{numpages}{25}~pages.
\newblock


\bibitem[Group(2023)]%
        {rust_lang_unsafe_code_guidelines_2019}
\bibfield{author}{\bibinfo{person}{Unsafe Code Guidelines~Working Group}.} \bibinfo{year}{2023}\natexlab{}.
\newblock \bibinfo{title}{Structs and Tuples - Memory Layout - Unsafe Code Guidelines}.
\newblock
\newblock
\urldef\tempurl%
\url{https://github.com/rust-lang/unsafe-code-guidelines/blob/50f8ff4b6892f98740de3b375e4d4bda10b9da9f/reference/src/layout/structs-and-tuples.md}
\showURL{%
\tempurl}
\newblock
\shownote{Accessed: Nov. 16 2019}.


\bibitem[Ho and Protzenko(2022)]%
        {aeneas-rust-verification-ho-2022}
\bibfield{author}{\bibinfo{person}{S. Ho} {and} \bibinfo{person}{J. Protzenko}.} \bibinfo{year}{2022}\natexlab{}.
\newblock \showarticletitle{{Aeneas: {Rust} Verification by Functional Translation}}.
\newblock \bibinfo{journal}{\emph{Proceedings of the ACM on Programming Languages}} \bibinfo{volume}{6}, \bibinfo{number}{ICFP} (\bibinfo{year}{2022}), \bibinfo{pages}{116:711--116:741}.
\newblock


\bibitem[Jacobs et~al\mbox{.}({[n.\,d.]})]%
        {quick-tour-verifast-jacobs-2010}
\bibfield{author}{\bibinfo{person}{B. Jacobs}, \bibinfo{person}{J. Smans}, {and} \bibinfo{person}{F. Piessens}.} \bibinfo{year}{[n.\,d.]}\natexlab{}.
\newblock \showarticletitle{A {Quick} {Tour} of the {VeriFast} {Program} {Verifier}}. In \bibinfo{booktitle}{\emph{Programming {Languages} and {Systems}}} \emph{(\bibinfo{series}{Lecture {Notes} in {Computer} {Science}})}. \bibinfo{address}{Berlin, Heidelberg}, \bibinfo{pages}{304--311}.
\newblock


\bibitem[Jung(2016)]%
        {scope-unsafe-jung-2016}
\bibfield{author}{\bibinfo{person}{R. Jung}.} \bibinfo{year}{2016}\natexlab{}.
\newblock \bibinfo{title}{The {{Scope}} of {{Unsafe}}}.
\newblock
\newblock
\urldef\tempurl%
\url{https://www.ralfj.de/blog/2016/01/09/the-scope-of-unsafe.html}
\showURL{%
\tempurl}
\newblock
\shownote{Accessed: March 3rd 2025}.


\bibitem[Jung(2018)]%
        {two-kinds-invariants-jung-2018}
\bibfield{author}{\bibinfo{person}{R. Jung}.} \bibinfo{year}{2018}\natexlab{}.
\newblock \bibinfo{title}{Two {Kinds} of {Invariants}: {Safety} and {Validity}}.
\newblock
\newblock
\urldef\tempurl%
\url{https://www.ralfj.de/blog/2018/08/22/two-kinds-of-invariants.html}
\showURL{%
\tempurl}
\newblock
\shownote{Accessed: June 19th 2023}.


\bibitem[Jung(2020)]%
        {understanding-evolving-rust-jung-2020}
\bibfield{author}{\bibinfo{person}{Ralf Jung}.} \bibinfo{year}{2020}\natexlab{}.
\newblock \emph{\bibinfo{title}{Understanding and evolving the {Rust} programming language}}.
\newblock {doctoralThesis}. \bibinfo{school}{Saarländische Universitäts- und Landesbibliothek}.
\newblock
\urldef\tempurl%
\url{https://doi.org/10.22028/D291-31946}
\showDOI{\tempurl}
\newblock
\shownote{Accepted: 2020-09-09T07:57:28Z}.


\bibitem[Jung et~al\mbox{.}(2019a)]%
        {stacked-borrows-aliasing-jung-2019}
\bibfield{author}{\bibinfo{person}{R. Jung}, \bibinfo{person}{H.-H. Dang}, \bibinfo{person}{J. Kang}, {and} \bibinfo{person}{D. Dreyer}.} \bibinfo{year}{2019}\natexlab{a}.
\newblock \showarticletitle{{Stacked Borrows: An Aliasing Model for {Rust}}}.
\newblock \bibinfo{journal}{\emph{Proceedings of the ACM on Programming Languages}} \bibinfo{volume}{4}, \bibinfo{number}{POPL} (\bibinfo{year}{2019}), \bibinfo{pages}{41:1--41:32}.
\newblock


\bibitem[Jung et~al\mbox{.}(2017)]%
        {rustbelt-securing-foundations-jung-2017}
\bibfield{author}{\bibinfo{person}{R. Jung}, \bibinfo{person}{J.-H. Jourdan}, \bibinfo{person}{R. Krebbers}, {and} \bibinfo{person}{D. Dreyer}.} \bibinfo{year}{2017}\natexlab{}.
\newblock \showarticletitle{{RustBelt: Securing the Foundations of the {Rust} Programming Language}}.
\newblock \bibinfo{journal}{\emph{Proceedings of the ACM on Programming Languages}} \bibinfo{volume}{2}, \bibinfo{number}{POPL} (\bibinfo{year}{2017}), \bibinfo{pages}{66:1--66:34}.
\newblock


\bibitem[Jung et~al\mbox{.}(2018)]%
        {iris-ground-modular-jung-2018}
\bibfield{author}{\bibinfo{person}{R. Jung}, \bibinfo{person}{R. Krebbers}, \bibinfo{person}{J.-H. Jourdan}, \bibinfo{person}{A. Bizjak}, \bibinfo{person}{L. Birkedal}, {and} \bibinfo{person}{D. Dreyer}.} \bibinfo{year}{2018}\natexlab{}.
\newblock \showarticletitle{{Iris from the Ground Up: {A} Modular Foundation for Higher-Order Concurrent Separation Logic}}.
\newblock \bibinfo{journal}{\emph{Journal of Functional Programming}}  \bibinfo{volume}{28} (\bibinfo{year}{2018}), \bibinfo{pages}{e20}.
\newblock


\bibitem[Jung et~al\mbox{.}(2019b)]%
        {future-ours-prophecy-jung-2019}
\bibfield{author}{\bibinfo{person}{R. Jung}, \bibinfo{person}{R. Lepigre}, \bibinfo{person}{G. Parthasarathy}, \bibinfo{person}{M. Rapoport}, \bibinfo{person}{A. Timany}, \bibinfo{person}{D. Dreyer}, {and} \bibinfo{person}{B. Jacobs}.} \bibinfo{year}{2019}\natexlab{b}.
\newblock \showarticletitle{{The Future is Ours: Prophecy Variables in Separation Logic}}.
\newblock \bibinfo{journal}{\emph{Proceedings of the ACM on Programming Languages}} \bibinfo{volume}{4}, \bibinfo{number}{POPL} (\bibinfo{year}{2019}), \bibinfo{pages}{45:1--45:32}.
\newblock


\bibitem[Lattuada et~al\mbox{.}(2023)]%
        {verus-verifying-rust-lattuada-2023}
\bibfield{author}{\bibinfo{person}{A. Lattuada}, \bibinfo{person}{T. Hance}, \bibinfo{person}{C. Cho}, \bibinfo{person}{M. Brun}, \bibinfo{person}{I. Subasinghe}, \bibinfo{person}{Y. Zhou}, \bibinfo{person}{J. Howell}, \bibinfo{person}{B. Parno}, {and} \bibinfo{person}{C. Hawblitzel}.} \bibinfo{year}{2023}\natexlab{}.
\newblock \showarticletitle{Verus: {Verifying} {Rust} {Programs} using {Linear} {Ghost} {Types}}.
\newblock \bibinfo{journal}{\emph{Proceedings of the ACM on Programming Languages}} \bibinfo{volume}{7}, \bibinfo{number}{OOPSLA1} (\bibinfo{year}{2023}), \bibinfo{pages}{85:286--85:315}.
\newblock


\bibitem[Lehmann et~al\mbox{.}(2022)]%
        {flux-liquid-types-lehmann-2022}
\bibfield{author}{\bibinfo{person}{N. Lehmann}, \bibinfo{person}{A. Geller}, \bibinfo{person}{N. Vazou}, {and} \bibinfo{person}{R. Jhala}.} \bibinfo{year}{2022}\natexlab{}.
\newblock \bibinfo{title}{Flux: {Liquid} {Types} for {Rust}}.
\newblock
\newblock
\urldef\tempurl%
\url{http://arxiv.org/abs/2207.04034}
\showURL{%
\tempurl}


\bibitem[Leroy et~al\mbox{.}(2012)]%
        {compcert-memory-model-leroy-2012}
\bibfield{author}{\bibinfo{person}{X. Leroy}, \bibinfo{person}{A.~W. Appel}, \bibinfo{person}{S. Blazy}, {and} \bibinfo{person}{G. Stewart}.} \bibinfo{year}{2012}\natexlab{}.
\newblock \emph{\bibinfo{title}{The {CompCert} {Memory} {Model}, {Version} 2}}.
\newblock Technical Report. \bibinfo{school}{Inria}.
\newblock
\urldef\tempurl%
\url{https://hal.inria.fr/hal-00703441}
\showURL{%
\tempurl}
\newblock
\shownote{Pages: 26}.


\bibitem[L{\"{o}}{\"{o}}w et~al\mbox{.}(2024)]%
        {cse}
\bibfield{author}{\bibinfo{person}{A. L{\"{o}}{\"{o}}w}, \bibinfo{person}{D. Nantes{-}Sobrinho}, \bibinfo{person}{S.{-}{\'{E}}. Ayoun}, \bibinfo{person}{C. Cronj{\"{a}}ger}, \bibinfo{person}{P. Maksimovi\'c}, {and} \bibinfo{person}{P. Gardner}.} \bibinfo{year}{2024}\natexlab{}.
\newblock \showarticletitle{{Compositional Symbolic Execution for Correctness and Incorrectness Reasoning}}. In \bibinfo{booktitle}{\emph{ECOOP'24}}. \bibinfo{pages}{25:1--25:28}.
\newblock
\urldef\tempurl%
\url{https://doi.org/10.4230/LIPIcs.ECOOP.2024.25}
\showDOI{\tempurl}


\bibitem[L{\"o}{\"o}w et~al\mbox{.}(2024)]%
        {matching-plans-frame-loow-2024}
\bibfield{author}{\bibinfo{person}{A. L{\"o}{\"o}w}, \bibinfo{person}{D. {Nantes-Sobrinho}}, \bibinfo{person}{S.-É. Ayoun}, \bibinfo{person}{P. Maksimovi{\'c}}, {and} \bibinfo{person}{P. Gardner}.} \bibinfo{year}{2024}\natexlab{}.
\newblock \showarticletitle{Matching {{Plans}} for {{Frame Inference}} in {{Compositional Reasoning}}}. In \bibinfo{booktitle}{\emph{ECOOP'24}}. \bibinfo{pages}{26:1--26:20}.
\newblock
\urldef\tempurl%
\url{https://doi.org/10.4230/LIPIcs.ECOOP.2024.26}
\showDOI{\tempurl}


\bibitem[Maksimović et~al\mbox{.}(2021)]%
        {gillian-part-ii-maksimovic-2021}
\bibfield{author}{\bibinfo{person}{P. Maksimović}, \bibinfo{person}{S.-É. Ayoun}, \bibinfo{person}{J. Fragoso~Santos}, {and} \bibinfo{person}{P. Gardner}.} \bibinfo{year}{2021}\natexlab{}.
\newblock \showarticletitle{Gillian, {Part} {II}: {Real}-{World} {Verification} for {JavaScript} and {C}}. In \bibinfo{booktitle}{\emph{CAV'21}}. \bibinfo{pages}{827--850}.
\newblock


\bibitem[Matsakis and Klock(2014)]%
        {rust-language-matsakis-2014}
\bibfield{author}{\bibinfo{person}{N.~D. Matsakis} {and} \bibinfo{person}{F.~S. Klock}.} \bibinfo{year}{2014}\natexlab{}.
\newblock \showarticletitle{{The Rust Language}}.
\newblock \bibinfo{journal}{\emph{ACM SIGAda Ada Letters}} \bibinfo{volume}{34}, \bibinfo{number}{3} (\bibinfo{year}{2014}), \bibinfo{pages}{103--104}.
\newblock


\bibitem[Matsushita et~al\mbox{.}(2022)]%
        {rusthornbelt-semantic-foundation-matsushita-2022}
\bibfield{author}{\bibinfo{person}{Y. Matsushita}, \bibinfo{person}{X. Denis}, \bibinfo{person}{J.-H. Jourdan}, {and} \bibinfo{person}{D. Dreyer}.} \bibinfo{year}{2022}\natexlab{}.
\newblock \showarticletitle{{RustHornBelt: A Semantic Foundation for Functional Verification of {Rust} Programs with Unsafe Code}}. In \bibinfo{booktitle}{\emph{PLDI'22}}. \bibinfo{pages}{841--856}.
\newblock


\bibitem[Matsushita et~al\mbox{.}(2021)]%
        {rusthorn-chcbased-verification-matsushita-2021}
\bibfield{author}{\bibinfo{person}{Y. Matsushita}, \bibinfo{person}{T. Tsukada}, {and} \bibinfo{person}{N. Kobayashi}.} \bibinfo{year}{2021}\natexlab{}.
\newblock \showarticletitle{{RustHorn}: {CHC}-based {Verification} for {Rust} {Programs}}.
\newblock \bibinfo{journal}{\emph{ACM Transactions on Programming Languages and Systems}} \bibinfo{volume}{43}, \bibinfo{number}{4} (\bibinfo{year}{2021}), \bibinfo{pages}{15:1--15:54}.
\newblock


\bibitem[Rahimi~Foroushaani and Jacobs(2022)]%
        {modular-formal-verification-foroushaani-2022}
\bibfield{author}{\bibinfo{person}{N. Rahimi~Foroushaani} {and} \bibinfo{person}{B. Jacobs}.} \bibinfo{year}{2022}\natexlab{}.
\newblock \bibinfo{title}{Modular {Formal} {Verification} of {Rust} {Programs} with {Unsafe} {Blocks}}.
\newblock
\newblock
\urldef\tempurl%
\url{http://arxiv.org/abs/2212.12976}
\showURL{%
\tempurl}


\bibitem[Sammler et~al\mbox{.}(2021)]%
        {refinedc-automating-foundational-sammler-2021}
\bibfield{author}{\bibinfo{person}{M. Sammler}, \bibinfo{person}{R. Lepigre}, \bibinfo{person}{R. Krebbers}, \bibinfo{person}{K. Memarian}, \bibinfo{person}{D. Dreyer}, {and} \bibinfo{person}{D. Garg}.} \bibinfo{year}{2021}\natexlab{}.
\newblock \showarticletitle{{RefinedC: Automating the Foundational Verification of C Code with Refined Ownership Types}}. In \bibinfo{booktitle}{\emph{PLDI'21}}. \bibinfo{pages}{158--174}.
\newblock


\bibitem[Team(2023a)]%
        {coq-proof-assistant--}
\bibfield{author}{\bibinfo{person}{The~Coq Team}.} \bibinfo{year}{2023}\natexlab{a}.
\newblock \bibinfo{title}{The {Coq} {Proof} {Assistant}}.
\newblock
\newblock
\urldef\tempurl%
\url{https://coq.inria.fr/}
\showURL{%
\tempurl}
\newblock
\shownote{Accessed: Nov. 16th 2023}.


\bibitem[Team(2023b)]%
        {how-open-source--2023}
\bibfield{author}{\bibinfo{person}{The~Kani Team}.} \bibinfo{year}{2023}\natexlab{b}.
\newblock \bibinfo{title}{How {Open} {Source} {Projects} are {Using} {Kani} to {Write} {Better} {Software} in {Rust} {\textbar} {AWS} {Open} {Source} {Blog}}.
\newblock
\newblock
\urldef\tempurl%
\url{https://aws.amazon.com/blogs/opensource/how-open-source-projects-are-using-kani-to-write-better-software-in-rust/}
\showURL{%
\tempurl}
\newblock
\shownote{Accessed: Nov. 13th 2023}.


\bibitem[Team(2023c)]%
        {rust-programming-language-therustteam-}
\bibfield{author}{\bibinfo{person}{The~Rust Team}.} \bibinfo{year}{2023}\natexlab{c}.
\newblock \bibinfo{title}{Rust {Programming} {Language}}.
\newblock
\newblock
\urldef\tempurl%
\url{https://www.rust-lang.org/}
\showURL{%
\tempurl}
\newblock
\shownote{Accessed: Nov. 16th 2023}.


\bibitem[Villani et~al\mbox{.}(2025)]%
        {tree-borrows-villani-2025}
\bibfield{author}{\bibinfo{person}{N. Villani}, \bibinfo{person}{J. Hostert}, \bibinfo{person}{D. Dreyer}, {and} \bibinfo{person}{R. Jung}.} \bibinfo{year}{2025}\natexlab{}.
\newblock \showarticletitle{Tree Borrows}.
\newblock \bibinfo{journal}{\emph{Proceedings of the ACM on Programming Languages}} \bibinfo{number}{PLDI} (\bibinfo{year}{2025}).
\newblock
\urldef\tempurl%
\url{https://doi.org/10.1145/3729291}
\showDOI{\tempurl}


\end{thebibliography}

\ifbool{extendedversion}{
\appendix

\section{Freezing existential variables}
\label{sec:appendix-freeze}

When performing borrow extraction in functions such as \rusti{LinkedList::first_mut}, one needs to \emph{freeze} existential variables introduced within the mutable borrow. The corresponding rule, \inferref{LftL-Bor-Exists} is given in Jung's thesis~\cite{understanding-evolving-rust-jung-2020}:
\[
\inferrule[LftL-Bor-Exists]{}{
  \fbw{\klft}(\exists x.~ P) \irupdate \exists x. \fbw{\klft} P
}
\]

For instance, consider the borrow $\fbw{\klft}(\exists y, z.~ x \mapsto y * y \mapsto z)$. If this borrow corresponds to a mutable reference, one could provide a sub-borrow $\fbw{\klft} (y \mapsto z)$. However, it is not possible to do so without saying that $y$ cannot change anymore. For this reason, one must start by first freezing~$y$, obtaining $\fbw{\klft}(\exists z.~ x \mapsto y * y \mapsto z)$. Then, one can split the borrow in two, thereby obtaining $\exists y.~ \fbw{\klft} (x \mapsto y) * \fbw{\klft} (\exists z.~ y \mapsto z)$, before discarding the first part.

The Gillian-Rust API provides the following macro to instantiate this rule for a given borrow and a given set of existentially quantified variables: 

\begin{lstlisting}
#[with_freeze_lemma(
  lemma_name = freeze_y,
  predicate_name = some_borrow_frozen,
  frozen_variables = [ y ]
)]
#[borrow]
fn some_borrow(x: *mut *mut i32) {
  gilsonite!(exists y: *mut i32, z: i32. x -> y * y -> z)
}
\end{lstlisting}

This \rusti{#[with_freeze_lemma(...)]} annotation generates two new items: a borrow where \rusti{y} is an input parameter instead of an existential variable; and a lemma that transforms the original borrow into the new one:

\begin{minipage}{.5\textwidth}
\begin{lstlisting}
#[borrow]
fn some_borrow_frozen(
  x: *mut *mut i32,
  y: *mut i32
) {
  gilsonite!(exists z: i32. x -> y * y -> z)
}
\end{lstlisting}
\end{minipage}%
\begin{minipage}{.5\textwidth}
\begin{lstlisting}
#[trusted]
#[lemma]
#[specification(
  requires { some_borrow(x) },
  exists y: *mut i32.
  ensures { some_borrow_frozen(x, y) }
)]
fn freeze_y(x: *mut *mut i32);
\end{lstlisting}
\end{minipage}

\section{Borrow extraction with prophecy variables}
\label{sec:appendix-extract}

Recall that, when proving type safety, the borrow-extract rule is as follows:
\[
\inferrule[Borrow-Extract]{
  \persistent{F}\\
  F * P \Rightarrow Q * (Q \wand P)
}{
  F * \alivetkn{\klft}_q * \fbw{\klft}P \irupdate \fbw{\klft}Q  * \alivetkn{\klft}_q
}
\]

We have proven this rule in Iris RustBelt development. Gillian-Rust provides a macro to instantiate a trusted lemma corresponding to the update in the conclusion of the rule, together with a proof obligation corresponding to the premise.

The premise says that, in the context of a persistent assertion $F$, assertion $Q$ can be derived from $P$, together with a wand $Q \wand P$, which ensures that, should the invariant $Q$ be restored, the assertion $P$ can be derived again. While this rule is sufficient to prove type safety of functions that perform borrow extraction, it is not enough to prove their functional correctness.

To prove functional correctness, one needs, in addition, to describe the relationship between the value contained in the original borrow, and that contained in the extracted borrow. This is done by introducing a function $f(a, b)$, where $a$ is the value of the original borrow, and $b$ is the value of the extracted borrow. The corresponding rule is the following:

\[
\inferrule[Borrow-Extract-Proph]{
  \persistent{F}\\ f(a, -)\text{ injective}\\\\
  F * P(a) \Rightarrow Q(b) * a = f(a, b) * (Q(b') \wand P(f(a, b')))
}{
  F * \alivetkn{\klft}_q * \fbw{\klft}(\exists a.~ P(a) * \pctrl{x}(a)) * \vobs{x}(a) \\\\
  \irupdate \fbw{\klft}(\exists b.~ Q(b) * \pctrl{y}(b)) * \vobs{y}(b) * \observt{\pret x = f(a, \pret y)} * \observt{a = f(a, b)} *
  \alivetkn{\klft}_q
}
\]

Let us walk through the rule step by step. The first premise is the same as in the previous rule. It allows us to perform extraction within the context of a persistent assertion $F$. For instance, when extracting the first node of a linked list, $F$ is the pure assertion that states that the list is not empty.

The second premise requires that the function $\lambda b.~ f(a, b)$ is injective. For instance, again in the case of extracting a borrow to the first element of a linked list, the function $f$ connects the representation of the list to the representation of the first element: $f(a, b) = b::(\mathtt{tail}~a)$. It is easy to check that this function is injective. 

The final premise is a generalisation of the premise of \textsc{Borrow-Extract}. It states that, if the invariant P holds for a value $a$, then the invariant $Q$ holds for a value $b$ such that $a = f(a, b)$, and for any $b'$, if $Q(b')$ holds, then it is possible to recover the invariant $P$ for the value $f(a, b')$. In the case of the linked list, this states that, given the entire linked list with representation $a$, one can extract a pointer to its first element with representation $b$, such that the entire list $a$ is the tail of $a$ with $b$ prepended. In addition, if the pointer to the first element is returned with representation $b'$, then the invariant for the entire linked-list is recovered with representation $f(a, b')$, that is, the tail of $a$ with $b'$ prepended.

The conclusion is an update which requires the context $F$ to hold, together with a lifetime token~$\alivetkn{\klft}_q$, and resource that has \emph{the same shape as the ownership predicate of a mutable reference}, with invariant $P$, prophecy variable $x$ and representation $a$. Such resource is usually either directly the ownership predicate of a mutable reference or a predicate obtained by freezing variable in the full borrow it contains. The update does not modify the lifetime token, and produces a new mutable-reference-like resource with invariant $Q$, prophecy variable $y$ and representation $b$. In addition, it \emph{partially resolves} the prophecy variable $x$, stating that the future value of $x$ (at the time when the borrow expires), denoted by $\pret{x}$, shall be $f(a, \pret{y})$, where $\pret{y}$ is the future value of $y$. Finally, it states the the current representation $a$ is equal to $f(a, b)$.

In the case of the linked list, the observations respectively state that the current value of the entire linked list is obtained by prepending the current value of the obtained pointer to the first element to the tail of the current list; and that the future value of the entire list is obtained by prepending the future value of the pointer to the first element to the tail of the current list. This is true since, when using the function \rusti{first_mut}, nothing other than the first element of the list can be modified until the borrow expires, as enforced by the borrow checker. 

Similarly to the case without prophecies, Gillian-Rust generates the obligation corresponding to the separation between the extracted resource and the magic wand. In addition, it generates a second obligation for the injectivity of the function $f(a, -)$, which is usually trivially discharged.

\section{EvenInt case study}
\label{sec:appendix-evenint}

\rusti{EvenInt} is a small structure used in the evaluation of RefinedRust, which we reuse as a basis for comparison. It contains a single value of type \rusti{i32}, for which the ownership invariant requires the value to be even. The applicable functions that we copy in our case study are:
\begin{itemize}[leftmargin=*]
\item \rusti{new} (unsafe), which receives an integer and returns an \rusti{EvenInt} without further~checks;
\item \rusti{new_2} (safe), which receives an integer, checks if it is even, and if it is not, adds or removes one to make it even, and then returns the corresponding \rusti{EvenInt};
\item \rusti{new_3} (safe), which receives an integer \rusti{i} and returns an \rusti{Option<EvenInt>}: \rusti{Some(i)} if \rusti{i} is even, and \rusti{None} otherwise;
\item \rusti{add} (unsafe), which increments the \rusti{EvenInt} value by one, breaking the soundness invariant; and
\item \rusti{add_two} (safe), which mutates an \rusti{EvenInt} in place, calling \rusti{add} twice.
\end{itemize}

We also import specifications from RefinedRust and rewrite them using Pearlite, the specification language of Creusot. Note that the specifications for \rusti{new_2} and \rusti{new_3} capture only type safety, whereas the specification of \rusti{add_two} guarantees both type safety and the simple functional correctness property that the value of the \rusti{EvenInt} object is incremented by two.

The total verification time of Gillian-Rust for the \rusti{EvenInt} study is \textbf{0.04s}, several orders of magnitude faster than the \textbf{4m36s} of RefinedRust. Furthermore, Gillian-Rust requires fewer specifications, as we can omit the specifications of the auxiliary internally unsafe functions \rusti{new} and \rusti{add}. We note, however, that one could write their specifications in Gillian-Rust and that doing so would not observably increase the verification time given the compositionality of Gillian.

In addition, in the \rusti{add_two} function, Gillian-Rust requires a single line of annotation to resolve the prophecy (the one in line~\ref{line:unique} of Figure~\ref{fig:impl-push-front}). In contrast, RefinedRust requires of the user to manually write a Rocq proof that if $i$ as in even integer, then $i + 1 + 1$ is still a even integer. The full code of the EvenInt case study for Gillian-Rust is provided below:

\begin{lstlisting}
struct EvenInt {
    num: i32,
}

impl Ownable for EvenInt {
    type RepresentationTy = i32;
    #[predicate]
    fn own(self, model: i32) {
        assertion!((self == EvenInt { num: model }) * (model % 2 == 0));
    }
}

impl EvenInt {
    #[creusillian::ensures(true)]
    pub fn new_2(x: i32) -> Self {
        if x % 2 == 0 {
            Self { num: x }
        } else {
            if x < 1000 {
                Self { num: x + 1 }
            } else {
                Self { num: x - 1 }
            }
        }
    }

    pub unsafe fn new(x: i32) -> Self {
        Self { num: x }
    }

    #[creusillian::ensures(true)]
    pub fn new_3(x: i32) -> Option<Self> {
        if x % 2 == 0 {
            let y = unsafe { Self::new(x) };
            Some(y)
        } else {
            None
        }
    }

    unsafe fn add(&mut self) {
        self.num += 1;
    }

    #[creusillian::ensures(true)]
    pub fn test(&mut self) {
        if self.num % 2 != 0 {
            panic!()
        }
    }

    #[creusillian::requires((*self@) <= i32::MAX@ - 2)]
    #[creusillian::ensures((^self@) == (*self@) + 2)]
    pub fn add_two(&mut self) {
        self.num;

        unsafe {
            self.add();
            self.add();
        }
        mutref_auto_resolve!(self);
    }
}
\end{lstlisting}
}{}

\end{document}